\documentclass[epj,nopacs]{svjour}
\usepackage{graphicx}
\usepackage{xspace}
\usepackage{amssymb}
\usepackage{amsmath}
\usepackage{xcolor}
\usepackage[colorlinks=true, allcolors=blue]{hyperref}

\newcommand{\pyt}{\textsc{Pythia}\xspace} 
\newcommand{\ang}{\textsc{Angantyr}\xspace} 
\newcommand{\cas}{\textsc{Cascade}\xspace} 

\def\AA{\ensuremath{AA}}
\def\NN{\ensuremath{NN}}
\def\pA{\ensuremath{pA}}
\def\hA{\ensuremath{hA}}
\def\hN{\ensuremath{hN}}
\def\pp{\ensuremath{pp}}
\def\pbarp{\ensuremath{p\overline{p}}}
\def\N14{\ensuremath{^{14}}\mbox{N}}
\def\O16{\ensuremath{^{16}}\mbox{O}}
\def\Ar40{\ensuremath{^{40}}\mbox{Ar}}
\def\Fe56{\ensuremath{^{56}}\mbox{Fe}}

\def\llangle{\left\langle}
\def\rrangle{\right\rangle}

\begin{document}
\title{Cosmic Ray Simulation with PYTHIA}
\author{Leif L\"onnblad \and Torbj\"orn Sj\"ostrand}
\institute{Division of Particle and Nuclear Physics,
Department of  Physics, Lund University, Lund, Sweden}
\date{Received: date / Revised version: date}
\abstract{ We present recent developments in \pyt for the modelling of
  hadronic cascades in a medium. Several improvements have been made
  in the Angantyr model for collisions with nuclei, especially in the
  limit of low collision energies, allowing it to be used throughout
  the hadronic cascades. Also the simplified nuclear model in the
  \textsc{PythiaCascade} module has been updated. We find that the two
  models give consistent results for cosmic ray air showers initiated
  by both high energy protons and nuclei.
} %end of abstract
\maketitle

\section{Introduction}
\label{intro}

High-energy cosmic rays can provide information on the most violent 
processes in the Universe. The flux is rapidly falling with energy,
however, so vast experimental detectors are necessary to study the 
extreme high-energy tail of events. Examples of such detectors are
Auger \cite{PierreAuger:2015eyc} for incident protons and nuclei,
and IceCube \cite{IceCube:2016zyt} for neutrinos. The incident
particle is not visible as such, but is detected by the interactions
it induces in the medium. The products of the primary collision
can interact in their turn, and so on, giving rise to a cascade of
particle production. It contains two main components, the hadronic
and the electromagnetic ones. The latter can arise as a by-product
of the former, e.g.\ when $\pi^0$ mesons decay to photons.

In this article we will study the hadronic cascade evolution in the
atmosphere. Core here is the modelling of a collision between a
single incoming hadron or nucleus and an atmospheric nucleus.
A number of different event generators have been developed to
describe such collisions \cite{Albrecht:2021cxw,Albrecht:2025kbb}, 
notably \textsc{DPMJet} \cite{Roesler:2000he},
\textsc{EPOS} \cite{Pierog:2013ria,Pierog:2023ahq,Werner:2023jps},
\textsc{QGSJet} \cite{Ostapchenko:2010vb,Ostapchenko:2024myl},
\textsc{Sibyll} \cite{Ahn:2009wx,Riehn:2019jet,Riehn:2023wdi}, and
\textsc{UrQMD} \cite{Bass:1998ca,Bleicher:1999xi}, 
each with several versions introduced over the years. In principle
these codes could also be used for the modelling of minimum-bias QCD
events at the LHC, but not for other LHC physics studies. Instead 
modelling of LHC \pp\ collisions is dominated by the three
general-purpose programs \pyt \cite{Bierlich:2022pfr}
\textsc{Herwig} \cite{Bewick:2023tfi}, and 
\textsc{Sherpa} \cite{Sherpa:2024mfk}. The one area of some overlap
at the LHC is heavy-ion collisions, \AA\ and \pA, where almost all
studies are of a QCD character. Thus the cosmic-ray class of generators
has also seen a fair amount of use, notably EPOS and \textsc{UrQMD},
as has \pyt, the only heavy-ion-enabled general-purpose generator,
with its \ang module \cite{Bierlich:2018xfw}.

It is unfortunate that the collider and cosmic-ray generator classes
are so separate from each other, when so many of the physics challenges
are common. Ideally one should strive to increase the overlap and
thereby the cross-fertilization, but from the collider side there are
some issues to overcome. For the simulation of cosmic ray cascades
it is not sufficient to handle \pp\, \pA\ and \AA\ interactions,
but also those of any secondary hadron sufficiently long-lived to
have time to interact before it decays or hits the ground: pions,
Kaons, Lambdas, Sigmas, and so on. Such capability was only added
to \pyt four years ago \cite{Sjostrand:2021dal}. Another issue is
that the tracking of an atmospheric cascade requires the rapid
switching between different beams, targets and collision energies.
This has not been possible within the \ang framework, so instead
the \textsc{PythiaCascade} --- henceforth \cas for simplicty ---
add-on was introduced as a simpler first approximation,
where such switching was possible. Notably, the full \ang
geometry setup was replaced by a simple parametrization of the
number of wounded nucleons. The \cas code does not handle incoming
nuclei, but in the field a standard trick has been to replace an A
nucleus beam with $A$ independent protons and neutrons, sharing the
full beam energy evenly \cite{Engel:1992vf,Ulrich:2010rg}.

In this article we present further development work, whereby \ang
now has been expanded so that it can handle the variability already
discussed above, with full nuclear geometry. This would make \cas
obsolete, except that the nuclear geometry shortcut still makes for
faster execution. It has also been an important reference in the
development and validation of the \ang extensions. One outcome is
that the new \ang setup clarifies the distinction between elastic
and inelastic cross sections, which in its turn paves the way
for some adjustments to \cas.

We implement a simple atmospheric model for some standalone studies
and comparisons presented in this article. The intention, however,
is that \pyt should act as a valid hadronic-interaction plugin for
the \textsc{Corsika}~8 tracking code \cite{CORSIKA:2023jyz},
and that is also now possible \cite{Reininghaus:2023ctx,Gaudu:2024rsq}.
In addition, already from its beginning \textsc{Corsika}~8 was
set up with \pyt~8 as its preferred particle decay engine, where it
helps that both codes are written in C++, unlike most cosmic-ray
generators.

The new capabilities\footnote{The new features presented in this
  paper will be available in \pyt version 8.317 and onwards.}
also are relevant for other cascade tracking
frameworks, notably the \textsc{Geant} detector simulation program
\cite{GEANT4:2002zbu} that is used extensively e.g.\ at LHC. Since
decades the earlier \pyt/\textsc{JetSet} \cite{Sjostrand:2006za} and
\textsc{Fritiof} \cite{Andersson:1986gw,Nilsson-Almqvist:1986ast}
Fortran codes have here been employed for hadronic interactions, but
ongoing work will make it possible to replace that with the new code
described in this article. In the future we also see applications to
hadronic cascades initiated by an incoming neutrino or photon, where it is
``only'' necessary to set up the primary interaction differently from
the setup studied here.

The recent and new developments that has enabled \pyt usage for
cosmic ray studies are introduced in Section~2. Thereafter Section~3
presents some toy studies, wherein the \ang and \cas setups are
compared. Finally Section~4 provides a summary and outlook.

\section{The new models}

The \pyt generator consists of several physics components,
to describe hard interactions, initial- and final-state parton
showers, matching and merging between interactions and showers,
multiparton interactions, beam remnants, colour reconnection,
fragmentation into hadrons, decays of unstable such, and more.
These components are described in the  \pyt~8.3 guide
\cite{Bierlich:2022pfr}, to which we refer.
Instead only extensions specific to cosmic-ray applications
are described here, first those that are generic enough to
become parts of the basic \pyt machinery, and afterwards those
specific to \ang and \cas, respectively.

\subsection{New core capabilities}

\pyt implements several models for \pp\ and \pbarp\ cross
sections \cite{Rasmussen:2018dgo}, with the default based on the
Donnachie--Landshoff (DL) approach \cite{Donnachie:1992ny}. In it,
the total cross section between two hadrons $A$ and $B$ is written
as the combination of a pomeron and a reggeon term:
\begin{equation}
\sigma_{\mathrm{tot}}^{AB} = X^{AB} s^{\epsilon} + Y^{AB} s^{-\eta}~,
\end{equation}
where $s$ is the squared total CM energy, and a fit gives
$\epsilon = 0.0808$ and $\eta = 0.4525$. The $X$ factor is
assumed to be the same for $\overline{A}B$ as for $AB$, while
the $Y$ factor is not. The DL article only considered a few
cross sections, but the approach was extended by Schuler and
Sj\"ostrand (SaS) \cite{Schuler:1993wr,Schuler:1996en}.
Thus $X$ and $Y$ coefficients have existed for $\pi^{\pm}p$,
$K^{\pm}p$, $\gamma p$, $\gamma\gamma$, and separately for
the $\rho^0 / \omega / \phi / J/\psi$ vector meson components
of the $\gamma$. This list has now been further expanded with
a number of other hadrons incident on a $p$
or a $\overline{p}$ \cite{Sjostrand:2021dal}:
$K^0$, $\eta$, $\eta'$, $D^{+,0}$, $D_s^+$, $B^{+,0}$, $B_s^0$,
$B_c^+$, $\Upsilon$, $\Lambda$, $\Xi$, $\Omega$, $\Lambda_c$,
$\Xi_c$, $\Omega_c$, $\Lambda_b$, $\Xi_b$ and $\Omega_b$.
Baryons that only differ by the relative composition of $u$ and
$d$ quarks or by their spin state are assumed equivalent, e.g.\
$\Lambda = \Sigma^+ = \Sigma^0 = \Sigma^-$.  The list 
includes hadrons sufficiently long-lived to interact
before they decay, some marginally so in a dilute atmosphere, but
with better chance in the passage through solid matter.

Since there does not exist data for most of these cross sections,
the Additive Quark Model (AQM) \cite{Levin:1965mi,Lipkin:1973nt}
is used, in which the pomeron $X$ term is assumed to be proportional
to the number of valence quarks, $X^{AB} \propto n_q^A n_q^B$,
extended with  a suppression for heavier quarks inversely proportional
to the respective constituent quark mass
\begin{equation}
n_{q,\mathrm{AQM}} = n_u + n_d + 0.6 \, n_s + 0.2 \, n_c + 0.07 \, n_b~. 
\end{equation}
The pattern of the $Y$ factors is less trivial to extend to new
unmeasured processes, but various considerations are used for educated
guesses, such as scaling by the number of light $u/d$ quarks that
can be exchanged between $A$ and $B$. In this spirit, neutron targets
are assumed to give the same cross sections as proton ones, and we will
use $N$ to denote a generic nucleon, $p$ or $n$, inside an equally
generic nucleus $A$.

One consequence of the DL ansatz is that cross section ratios, such as
$\sigma_{\mathrm{tot}}^{\pi p} / \sigma_{\mathrm{tot}}^{pp} \approx 2/3$,
remain fixed at high energies. This is different from most
other cosmic-ray generators, where it is assumed that all cross
sections eventually converge, i.e.\
$\sigma_{\mathrm{tot}}^{\pi p} / \sigma_{\mathrm{tot}}^{\pp} \to 1$
\cite{Reininghaus:2023ctx}, based on the assumption of a universal
saturation of the small-$x$ gluon cloud. We do not know which is
correct, if any, so it is necessary to keep an open mind.

The extensions to elastic and diffractive cross sections introduce
further complications, and become too detailed to cover in full here,
but the basic framework is the SaS one. The elastic slope
$B_{\mathrm{el}}^{AB}$ here is a combination of hadron-specific fixed
terms and a common $s^{\epsilon}$ pomeron exchange term, and from it
the optical theorem gives the elastic cross section. The choice of an
$s^{\epsilon}$ term rather than the conventional $\ln(s)$ term in
Regge theory, one ensures that $\sigma_{\mathrm{el}}(s)$ never exceeds
$\sigma_{\mathrm{tot}}(s)$.  For single and double diffraction again a
modified Regge ansatz is used, which approximately preserves a
diffractive mass spectrum shape like $\mathrm{d} m^2 / m^2$, with a
low-mass enhancement.  The differential cross sections are integrated
numerically, and their energy dependence parametrized to provide a
quick selection of process type to generate in a collision. The
inelastic nondiffractive cross section, which is the largest
component, is defined by what is left after elastic and diffractive
cross sections have been subtracted from the total one.

At energies close to threshold also other processes contribute,
such as annihilation or the formation of an intermediate resonance.   
A framework for low-energy total and partial cross sections was
introduced to \pyt in Ref.~\cite{Sjostrand:2020gyg}, and is used here.
It is smoothly matched to the higher-energy behaviour described
above.

Once a collision has been deemed to occur, a (mainly hadronic)
final state has to be generated. Elastic scatterings are trivial,
given the assumed $B_{\mathrm{el}}^{AB}$ above. For inelastic
nondiffractive events a simple nonperturbative handling is used
for CM energies up to 10 GeV, and thereafter the perturbative
multiparton interaction (MPI) machinery gradually takes over.

In the nonperturbative handling, the interaction is assumed driven by
the exchange of a gluon, of vanishing momentum but turning the colliding
particles into colour octets. (This can be viewed as the
$n_{\mathrm{MPI}} = 1, p_{\perp} \to 0$ limit of the MPI framework.)
Considering only the valence content,
a hadron thus is composed of a triplet quark, and a antitriplet
antiquark or diquark for a meson or baryon respectively. Thereby
two standard Lund strings are stretched out, each between the triplet
of one hadron and the antitriplet of the other. These fragment
independently of each other, to give the primary hadron products.

In the perturbative machinery, the MPI framework is driven by the
perturbative interactions, such as $qq \to qq$, $qg \to qg$ and
$gg \to gg$. To calculate their cross sections it is necessary
to have access to parton distribution functions (PDFs) for all hadrons
that may interact. Apart from the $p/n$, only little is know about
the $\pi^{\pm}$, even less about $K^{\pm}$, and the rest is silence.
Therefore the new SU21 family of PDF sets \cite{Sjostrand:2021dal}
is introduced for $p$, $\pi$, $K$, $\eta$, $\phi$, $D$,
$D_s$, $J/\psi$, $B$, $B_s$, $B_c$, $\Upsilon$, $\Sigma$,
$\Xi$, $\Omega$, $\Sigma_c$, $\Xi_c$, $\Omega_c$,
$\Sigma_b$, $\Xi_b$ and $\Omega_b$, where the first two
are included for completeness but by default replaced by existing
tunes in the literature.
Inspiration comes from the leading-order GRS99 $\pi^+$ set
\cite{Gluck:1999xe}, which is dynamically generated from the low
starting scale $Q_0^2 = 0.26$ GeV$^2$. At this scale, the SU21
valence quarks are parametrized to be of the form $N x^a (1-x)^b$.
Heavier quarks are assumed to take a larger fraction of the hadron
momentum than $u$ or $d$ ones (or than gluons), so as roughly to
make all valence partons of a hadron have the same average velocity,
if compared in terms of constituent masses.The $u$ and $d$
valence quark distributions are assumed equal within a set, and the
same for all hadrons with a common heavier quark content, limiting
the number of separate sets needed. The starting PDFs are evolved
with QCDNUM \cite{Botje:2010ay} and the resulting grid files are
stored in \pyt.

Apart from the new PDFs, and the need to handle mass effects for the
heavy $c$ and $b$ quarks, the MPI machinery is the same as for \pp.
Specifically it is assumed that $p_{\perp 0}$, the lower
regularization scale of the divergent $2 \to 2$ cross section, has the
same energy dependence for all hadrons, which does not have to be the
case. The $p_{\perp 0}$ choice,
together with the PDFs, regulates the integrated MPI cross
section. The average number of MPIs is then given by the ratio of
the MPI cross section and the total nondiffractive one. This average
feeds into the final-state multiplicity of a collision.

A technical complication is that, for each incoming hadron type and
each collision energy, it is necessary to have an analytical
overestimate of the differential cross section in a given
$(x_1, x_2, p_{\perp})$ phase space point, that can be used for Monte Carlo 
hit-and-miss selection down to the correct cross section. Also
the integrated cross section above a given $p_{\perp}$ should be available
for use in a Sudakov-style rejection factor. The required
initialization may only take a few seconds, but that could
easily be a factor 1000 slower than it takes to generate an event,
so is not an option inside a cascade, where beam particles and energies
change all the time. Instead results are tabulated
for each of the 21 allowed incoming hadrons, and for each of them
in a grid of energies, starting at 10~GeV and then logarithmically
distributed upwards. For each new collision in a cascade, relevant
numbers are found by energy interpolation in the respective hadron case. 
The full tables take several minutes to generate, but can be stored on
an external file and be reused so long as none of the parameters
affecting the MPI cross sections are changed. This allows for a
significant speed-up of the cascade evolution. 

In the spirit of the Ingelman--Schlein approach \cite{Ingelman:1984ns},
a (single) diffractive subsystem can be viewed as a pomeron--hadron
subcollision, where the pomeron is a composite hadron-like object,
with a two-gluon valence-like content. Therefore $Q^2$-dependent PDFs
can be defined, and that leads to MPIs. Once the cross section
machinery has selected a diffractive mass, the MPI activity
only depends on this mass and not on the full energy of the collision. 
Therefore a similar pretabulation strategy can be used as for
nondiffractive events, over a grid in mass rather than in CM energy.
All 21 hadrons have to be tabulated separately when the incoming
hadron is diffractively scattered, given the difference in PDFs.
Diffraction on the proton side only has to be tabulated once.
For double diffraction the two systems can be handled separately,
and so no further tabulation is required. Central diffraction,
finally, can be viwed as a pomeron--pomeron collision, and only
introduces one further case.

MPIs take out partons from the incoming hadrons, leaving behind a
set of valence and sea remnant quarks that have to be hooked up with the
rest of the outgoing partons into singlets, that then fragment. 
The handling of the beam remnants has a direct impact on hadron
production in the forward region, which is crucial in the modelling
of atmospheric cascades, since the production of a single hadron
that takes almost all the incident energy (low inelasticity) gives
a slower evolution of the cascade than one where the original energy
is shared more equally between many hadrons (high inelasticity).

Indications are that, in \pp\ collisions, the \pyt forward $p/n$
energy spectrum is too soft and the forward $\gamma/\pi^0$ too hard,
in spite of an energy sharing prescription where any diquark takes
the bigger chunk of the remnant momentum. To address this issue, in
Ref.~\cite{Fieg:2023kld} two main optional changes were introduced
to make the baryon harder, i.e\ take more momentum.
One is to reduce the so-called popcorn mechanism for a diquark at the
endpoint of a string, which allows the diquark to disconnect so that
a meson can be produced at the end of the string, with the baryon
produced only in the second step. The other is to modify the Lund
symmetric fragmentation for an endpoint diquark. The natural choice
would be to use a modified $a$ parameter for diquarks, but this turned
out to give only small effects, so both modified $a$ and $b$ values
were used. Recently, however, a bug was found in the handling of
effects when a different $a$ is picked for diquarks, as described
in the \pyt~8.315 release documentation. Preliminarily it seems that
this fix obviates the need to modify $b$, but more detailed studies
are needed. For the moment being therefore we content ourselves with
the pre-8.315 approach. It remains to be studied whether also the
forward region of meson beams needs to be modified.

\subsection{The \ang framework}

\ang relies on the Good--Walker formalism for the Glauber calculation
of cross sections and for generating events. In this way fluctuations
in the nucleon--nucleon (\NN) cross sections can be taken into
account, in the spirit of Gribov's corrections to the original Glabuer
formalism \cite{Gribov:1968jf}. This means that \ang describes, not
only \textit{if} a nucleon in the projectile nucleus interacts with
one in the target nucleus, but also \textit{how} they interact. The
procedure is described in detail in \cite{Bierlich:2018xfw}, and here
we will only emphasise the main points.

The fluctuations in the nucleon wavefunction are parameterised in
terms of an interaction radius, that also influences the overall
elastic amplitude. By default, the this radius if allowed to fluctuate
according to a Gamma distribution,
\begin{equation}
  \label{eq:gamma-dist}
  P(r)=\frac{r^{k-1}e^{-r/r_0}}{\Gamma(k)r_0^k},
\end{equation}
where $k$ and $R_0$ are parameters, and $\Gamma(k)$ is the standard
Gamma function.  The elastic amplitude (which is approximated to be
purely imaginary, with $T=\text{Im}(A_\text{EL})$) is then assumed to
given by
\begin{equation}
  \label{eq:1}
T(b,\sigma)=T_0(\sigma)\,\,\Theta\!\left(\sqrt{\frac{\sigma}{2\pi T_0(\sigma)}}-b\right),
\end{equation}
where $\sigma=(r_p+r_t)^2$ is the squared sum of the projectile and
target nucleon radii, and
\begin{equation}
  \label{eq:tzero}
  T_0(\sigma) = \left(1-\exp\left(-\frac{\sigma_T}{\pi\sigma}\right)\right)^\alpha.
\end{equation}
Here, $\sigma_T$ and $\alpha$ are parameters which, together with the
constants in the Gamma distribution can be fitted to reproduce
semi-inclusive \NN\ cross sections. From the Good--Walker formalism we
get the following expressions for the total, non-diffractive
inelastic, elastic, and diffractive excitation of the projectile,
target and double diffractive excitation:
\begin{eqnarray}
  \label{eq:nnxsecs}
  d\sigma^{\NN}_\text{tot}/d^2b &=& \llangle 2T(b)\rrangle_{p,t}\nonumber\\
  d\sigma^{\NN}_\text{ND}/d^2b &=& \llangle 2T(b)-
                          T^2(b)\rrangle_{p,t},\nonumber\\
  d\sigma^{\NN}_\text{EL}/d^2b &=& \llangle T(b)\rrangle^2_{p,t}\nonumber\\
  d\sigma^{\NN}_\text{pex}/d^2b &=& \llangle\llangle T(b)\rrangle_t^2\rrangle_p -
                          \llangle T(b)\rrangle^2_{p,t}\nonumber\\
  d\sigma^{\NN}_\text{tex}/d^2b &=& \llangle\llangle T(b)\rrangle_p^2\rrangle_t -
                          \llangle T(b)\rrangle^2_{p,t}\nonumber\\
  d\sigma^{\NN}_\text{Dex}/d^2b &=& \llangle T^2(b)\rrangle_{p,t} +
                         \llangle T(b)\rrangle^2_{p,t}\nonumber\\ -
                         &&\llangle\llangle T(b)\rrangle_p^2\rrangle_t -
                         \llangle\llangle T(b)\rrangle_t^2\rrangle_p.
\end{eqnarray}
Here $\llangle\cdots\rrangle_p$ and $\llangle\cdots\rrangle_t$ are
averages over projectile and target radius respectively, and we note
how the diffractive excitations are intimately connected to the
fluctuations in the wavefunctions.

From the factorisation of the \NN\ $S$-matrices, $S_{pt}=1-T_{pt}$, we
can write down the corresponding nucleus--nucleus (\AA) cross sections
using the combined elastic amplitude
\begin{equation}
  \label{eq:TAA}
  T^{\AA}(b)=1-\prod_{p}\prod_{t}\left(1-T^{\NN}(b_{pt},r_pr_t)\right),
\end{equation}
where $\pmb{b}_{pt}=\pmb{b}_{p}+\pmb{b}-\pmb{b}_{t}$ is the relative
impact parameter separation between a projectile and target nucleon,
assuming an overall \AA\ impact parameter, $\pmb{b}$ and individual
impact parameters, $\pmb{b}_{p/t}$, relative to the respective nuclei.

It should be noted that, although \ang uses the expressions above
to calculate the \NN\ and \AA\ semi-inclusive cross sections, the
generated events are not exactly distributed in the same
way. Considering the amplitude, $T$, always being positive and less
than unity, it is tempting to use the different $d\sigma^{\NN}/d^2b$
for a given $b$, $r_p$ and$ r_t$, as a probabilities for the different
processes to occur, but since most of them includes squared averages,
that is not generally possible. It does work for the non-diffractive
interaction, however. Here we can generate an impact parameter, an
$r_p$ and an $r_t$ for the nucleons, and treat it as a probability for
a non-diffractive scattering. Summing over all events will then
correspond to an integration over impact parameter and averaging over
projectile and target states and will give the correct non-diffractive
cross section. But this is not possible for the other semi-inclusive
cross sections. And even if we could, we note that for the ``black''
case of $T=1$ the total ``probability'' would be 2, which is clearly
nonsense.

The procedure implemented in \ang is to generate positions of each
nucleon according to the so-called
GLISSANDO~\cite{Broniowski:2007nz,Rybczynski:2013yba}
parametrrisation,\footnote{The GLISSANDO parameterisation used is
  formally only valid for $A>16$, but \ang uses it also for lighter
  nuclei by default. Options for lighter nuclei are available.} an
overall impact parameter,\footnote{By default \ang samples the whole
  impact parameter space using a Gaussian weighting, giving rise to
  weighted events, but an option to only sample within a limited
  region giving unweighted events is available} and a radius for each
nucleon. In addition an auxiliary radius, $r'_{p/t}$, is sampled for
each nucleon, in order to also sample the fluctuations in radius. In
this way we have for each \NN\ (and \AA) collision four statistically
equivalent samples, and the general idea is to ``shuffle
probabilities'' between these to get the probability for a given
process to happen a particular event.

To illustrate this we try to get the probability for a single
diffractive target excitation as
\begin{equation}
  \label{eq:4}
  P_{t\text{EX}}=T(b,r_p,r_t)T(b,r_p,r'_t)-T(b,r_p,r_t)T(b,r'_p,r'_t),
\end{equation}
which, when summed over many events will result in the correct cross
section on the form
$\langle\langle T\rangle_p^2\rangle_t-\langle\langle
T\rangle_{pt}^2\rangle $ in eq.~(\ref{eq:nnxsecs}). This probability
can, however, become negative, which we solve by looking also at the
other combinations with auxiliary states, to write the average
probability
\begin{eqnarray}
  \label{eq:avnntex}
  P_{\text{tex}}&=&\frac{1}{4}\left(T(b,r_p,r_t)T(b,r'_p,r_t)-T(b,r_p,r_t)T(b,r'_p,r'_t)\right.+\nonumber\\
  &&\phantom{\frac{1}{4}..}T(b,r'_p,r_t)T(b,r_p,r_t)-T(b,r'_p,r_t)T(b,r_p,r'_t)\nonumber\\
  &&\phantom{\frac{1}{4}..}T(b,r_p,r'_t)T(b,r'_p,r'_t)-T(b,r_p,r'_t)T(b,r'_p,r_t)\nonumber\\
  &&\phantom{\frac{1}{4},}\left.T(b,r'_p,r'_t)T(b,r_p,r'_t)-T(b,r'_p,r'_t)T(b,r_p,r_t)\right)\nonumber\\
  &=&\frac{1}{2}\left(T(b,r_p,r_t)-T(b,r_p,r'_t)\right)\times\nonumber\\
&&\phantom{\frac{1}{4}}\left(T(b,r'_p,r_t)-T(b,r'_p,r'_t)\right).
\end{eqnarray}
This expression is always positive and less than unity as long as $T$
is a monotonic function in $r_t$, which is the case in
eq.~(\ref{eq:tzero}). We can therefor distribute this probability
between the four states and make sure it is always positive.

In fact, the form of $T$ we have chosen in eq.~(\ref{eq:tzero}) is
separately monotonic in $r_p$ and $r_t$, which allows us to do the
same for the other inelastic diffractive cross sections.\footnote{In
  the current version of \pyt this procedure only gave the correct
  ``wounded'' cross sections, i.e., the sum of ND, DD and $p$EX or
  $t$EX ones, but here we will use a better stategy where all
  inelastic cross sections come out right.} Unfortunaterly, this
procedure cannot be applied to give the correct elastic cross section.

After the Glauber simulation \ang has determined if and how any
combination of projectile and target nucleon will interact. Special
care must be taken in the case where, e.g., one projectile has
interacted with more than one target nucleon. The procedure to handle
this is to order all potential interactions according to their impact
parameter, $b_{pt}$, starting with the most central \NN\
sub-collision. If neither of the nucleons in such a \NN\ pair has
interacted before, the interaction is labelled primary, and a
corresponding sub-event is generated by the standard machinery in
\pyt. If both have interacted, the sub-collision is simply ignored,
while if only one has interacted, the interactions is labelled
secondary and treated as a diffractive excitation of the other
nucleon, according to a procedure described in detail in
\cite{Bierlich:2018xfw} and further refined in \cite{9131895}.

By default, the current \pyt/\ang version tries to fill in as much elastic
\NN\ scatterings as possible, but it will not give the correct
weight. This is a cause of some confusion for some users, as the
\texttt{Pythia::stat()} function, in the case of nuclear collisions,
will print out both the cross sections corresponding to the generated
events, and the cross section obtained from the Glauber calculation,
which do not match each other. This is expected for the elastic cross
section, but is also true for the other ones. This is because the
generated cross sections are listed according to the type of the
\textit{primary} \NN\ collision, while the Glauber calculation gives
the cross section for the according to the type of the full \AA\
collision. As an example of this, consider an \AA\ collision with only
one \NN\ single diffractive target excitation scattering, which is
then also the primary interaction, and the event will included as such
in the generated cross section in the listing. In the Glauber
calculation for the overall \AA\ cross sections, however, such an
event will be counted as a double diffraction event, as the
elastically scattered projectile nucleon will cause that nucleus to
break up as well.

To enable the generation of hadronic cascades using \ang, several
developments has been made. First of all we have introduced the
possibility of changing both the beam particles and the collision
energy during a run. Normally \ang needs a rather lengthy
initialisation procedure, involving refitting of the nucleon
fluctuations to the relevant beams and energies, but now we have
introduced a caching where the initialisation of a wide range of
energies and beams,can be done once and for all and save to disk, to
be read in future runs.

The procedure described so far is only suitable for high collision
energies. For energies below $\sim20$~GeV the fit to the
semi-inclusive cross sections becomes very poor, and the whole fitting
procedure becomes unstable. For lower energies \ang will instead use a
simplified Glauber model, where the kind of sub-collision is
determined only from the impact parameter, $b_{pt}$, Basically, if
$\pi b_{pt}^2<\sigma^{\NN}_\text{ND}$ the sub-collision is assumed to
be non-diffractive, otherwise if
$\pi b_{pt}^2<\sigma^{\NN}_\text{ND} + \sigma^{\NN}_\text{DEX}$ we get
a doubly diffractive excitation, and so on. In this way we also
include the additional possible interactions in \pyt's low energy
interaction machinery. This is a very naive model, which we hope to
improve on in the future.

For this paper we have also introduced an alternative procedure in
\ang to be used for hadronic cascades.\footnote{available in the
  setting \texttt{\ang:cascadeMode = on}.} In this mode, the total
generated cross section should be very close the total inelastic \AA\
cross section. Using the unweighted impact parameter sampling it is
then possible get an overestimated cross section (simply given by the
sampling area). This can then be used to randomly generate the
position of the next hadron collision in a medium. When calling
\texttt{Pythia::next()}, \ang will now only generate one impact
parameter point together with the radii and positions of all nucleons,
and return \texttt{false} if no collision was produced. The user
should then generate a new collision point, taking the previous failed
one as starting point, and try again, and so on. The advantage of this
is that there is no need for parameterising all cross sections for all
types of hadrons and nuclei, these are instead obtained
\textit{on-the-fly} through the Glauber procedure.

For the results presented in this paper have chosen to not include
elastic \NN\ scatterings at all in the generation. This means that
when we discuss the total inelastic \AA\ cross section, we are
referring to the part of the total cross section that involves
inelastic \NN\ scatterings. This is a reasonable approximation since
the momentum of an elastically scattered incident hadron is only
slightly changed and the slightly excited target nuclei is typically
not observable. Hence the effect on the overall hadronic cascade
should be very small.

For incident cosmic nuclei, such elastic scatterings may cause a
nucleus to break up and initiate separate cascades, so there may be
visible effects in this case. The modelling of nucleus remnants in
\ang is, however, very crude in general, and the effect of omitting
these elastic scatterings is expected to small compared to the large
uncertainties in the remnant modelling.

\subsection{The \cas framework}

Given the \ang limitations at the time, \cas was intended as a
simpler, more flexible alternative, where the machinery for 
hadron--nucleon (\hN) collisions of section 2.1 is
extended to apply to hadron--nucleus (\hA) ones. It is also
structured to rapidly provide the cross sections relevant for a mixed
medium, such as air or ice, and on demand to perform either a collision
or a decay.

Following recent modifications to \ang, and clarifications as to how
its event classification works, the original \cas behaviour has been 
updated for this article. We will begin this section with an overview
of the original \cas, and afterwards summarize those new updates.

In \ang the selection of nuclear geometry, i.e. the location of
the individual nucleons in a nucleus and the incoming hadron
impact parameter, takes a fair amount of time. The key information
that comes out of this modelling is the number of wounded nucleons,
i.e. the number of nucleons in the target that undergo a subcollision
with the hadron. This probability distribution turns out to be
approximately a geometrical series, and so is characterized uniquely
by its average number $\langle n_{\mathrm{wound}} \rangle$. For a given
nucleus $A$ this average depends on which incoming hadron type hits
with what energy, but only through the total $\hN (=hp)$ cross
section, as a measure of the size of the interaction region spanned by
the hadron in its passage through the nucleus. The average grows
approximately linearly with the \hN\ cross section, with some
corrections at small cross sections, so is easily parametrized.
Currently this has been done for fifteen common nuclei, from deuterium
to lead, with interpolation for intermediate ones.

The \hA\ cross section is now chosen to be
\begin{equation}
\sigma_{\mathrm{tot}}^{\hA} = \sigma_{\mathrm{tot}}^{\hN} \,
\frac{A}{\langle n_{\mathrm{wound}} \rangle}~.
\label{eq:sigmaTothA}
\end{equation}
so that the cross section for an individual nucleon to be hit
is the same whether it is bound inside a nucleus or free-floating. 

Thus the geometry handling is reduced to a simple scheme, starting out
from a target with $n_p = Z$ protons and $n_n = A - Z$ neutrons:
\begin{enumerate}
\item find the \hN\ total cross section;
\item turn this into $\langle n_{\mathrm{wound}} \rangle$, which gives
the geometric series ratio $r = 1 - 1 / \langle n_{\mathrm{wound}} \rangle$;
\item wound a proton or a neutron  in proportions $n_p : n_n$;
\item generate the relevant subcollision, and update $n_p$ and $n_n$;
\item break out of the loop with probability $1 - r$, or if there
are no nucleons left;
\item else loop to 3.
\end{enumerate}
The fate of the unwounded target nucleons is not addressed in this
scheme. Some of them may form a new nucleus, while others break free. 
In the fixed-target context, however, all will have small momenta,
and almost always stop before they reach the ground.

Besides the geometry handling, also the event shape needs to be simulated.
Each \hN\ collision here contributes to the particle production.
But each collision can take less of the incoming hadron momentum the
more collisions there are. This should be reflected in a softening of
the outgoing momentum spectrum for an increasing $n_{\mathrm{wound}}$,
and a less-than-linear increase of the total multiplicity. In the \ang
model each collision leads to stretched-out Lund strings from the
wounded nucleon towards the incoming hadron, e.g. as considered along
a rapidity axis. Only one collision gives strings stretched all the
way to the hadron, however, while the remaining have a topology more
similar to a diffractive system, where the strings end in a flavour-less
pomeron-like object that can be seen as having been emitted by the
incoming hadron.

The details are more complicated, and again a simpler solution was
created for \cas, with the intent to reproduce the overall pattern
of \ang final-state distributions. In it, the first collision
(in order of handling, with no collision time order implied) is
generated as a normal \hN\ one. The final-state particle with the
highest momentum (in the direction of the incoming hadron) is then
made projectile for the next subcollision, and so on. The choice of
flipping the projectile identity should be viewed as a technical trick
to ensure flavour conservation, not as a physical change of propagating
particle. Specifically, high-energy particles are produced only after
the whole $hA$ collision. A simple rule of thumb is that particles with
a momentum $p$ are produced a distance $\kappa p$ away from the collision
vertex, with the string tension $\kappa \approx 1$~GeV/fm. Therefore
the identity flip does not affect the original $hN$ cross section,
as reflected in the  $n_{\mathrm{wound}}$ distribution. For technical
reasons, however, the handling of MPI activity in an $hN$ collision
is based on the current $h$ particle rather than on the original one.
Since e.g.\ $\pi p$ and $p p$ give slightly different multiplicities
this is not fully consistent, but the
$\pi \leftrightarrow K \leftrightarrow p$ flipping inside an $hA$
collision reflects a similar flavour composition evolution going on
in the whole atmospheric cascade, so already small differences are
further diluted.
 
Remains to discuss the nature of each subcollision. Based on the
contemporaneous understanding of \ang cross sections, the first
collisions is assumed to be of any type, elastic or inelastic,
while any subsequent ones are always inelastic. More specifically,
a nondiffractive event is assumed to occur with probability 0.7,
with the remaining corresponding to single diffraction of the target
nucleon. This then roughly reproduces the \ang charged multiplicity
distribution and the $\mathrm{d} n_{\mathrm{charged}} / \mathrm{d} y$\phantom{9} shape.
For low collision energies, where \ang does not offer any guidance,
the standard \pyt mix of subprocesses is used.

\begin{figure}
\includegraphics[width=\columnwidth]{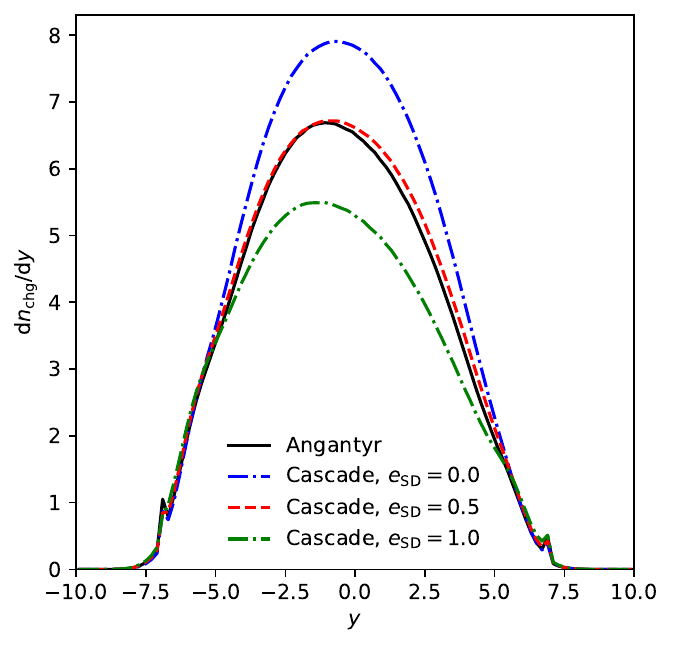}
\caption{Charged-particle rapidity distribution in $p\N14$ collisions
at 1~TeV proton--nucleon energy, comparing three values of the
$e_{\mathrm{SD}}$ parameter introduced in the text.}
\label{fig:enhanceSD}
\end{figure}

Since the 2021 introduction of the \cas code, and notably in the last
year, the generation and presentation of \ang cross sections
has evolved, as described above. One point of confusion has been that 
the total cross section includes both incoherent and coherent elastic
scattering, i.e. where the nucleus does or does not break up. The 
latter does not obey eq.~(\ref{eq:sigmaTothA}), but comes in 
addition. Furthermore, \ang cannot cleanly separate the two
contributions, and so only generates inelastic collisions. By contrast,
\cas was originally set up to generate the whole incoherent cross section,
including a first elastic scattering, but any subsequent interactions
was only inelastic to agree with \ang in this respect. Therefore \ang
has given a higher $\sigma_{\mathrm{tot}}^{hA}$ than \cas, but a lower 
$\sigma_{\mathrm{inel}}^{hA}$. Now elastic scatterings are removed altogether
from \cas, either as first or as subsequent scattering. Thus
eq.~(\ref{eq:sigmaTothA}) comes to be a statement about
$\sigma_{\mathrm{inel}}$ rather than $\sigma_{\mathrm{tot}}$,
and $n_{\mathrm{wound}}$ only counts inelastically wounded nucleons:
\begin{equation}
\sigma_{\mathrm{inel}}^{\hA} = \sigma_{\mathrm{inel}}^{\hN} \,
\frac{A}{\langle n_{\mathrm{wound}}^{\mathrm{inel}} \rangle}~.
\label{eq:sigmaInelhA}
\end{equation}
No longer having access to $\sigma_{\mathrm{tot}}^{hA}$ is not a
limitation in the atmospheric cascade community, where 
$\sigma_{\mathrm{inel}}^{hA}$ traditionally is used to compare generators.
$\ang$ bug fixes in the handling and bookkeeping of collisions has 
also led to increased  $\langle n_{\mathrm{wound}} \rangle$ by the order
of $\sim10$\%, and new parametrizations of the 
$\langle n_{\mathrm{wound}} \rangle (\sigma_{\mathrm{inel}}^{hA})$
have been made accordingly.

Also the nature of secondary collisions has been revisited. The removal
of elastic scatterings and other changes in cross section calculations,
the above-mentioned increase of $\langle n_{\mathrm{wound}} \rangle$,
and other minor changes, contribute to make the previous
subprocess mix obsolete. Furthermore, a richer set of diffractive and
nondiffractive topologies are allowed in \ang than assumed in the past
(or well documented), which also should be taken into account. Therefore
a new mix is introduced for the secondary collisions, where a fraction
$e_{\mathrm{SD}}$ is modelled as target-side diffraction, while the rest
is given as the standard composition of all inelastic collision types,
including a further small fraction of target-side diffraction. A value
of $e_{\mathrm{SD}} = 0.5$ gives fair agreement between \cas and \ang,
as seen in Fig.~\ref{fig:enhanceSD}, and is used as default. 
Nevertheless it is a number not carved in stone, but rather offers 
a key parameter that easily can be varied to explore uncertainties
in the atmospheric cascade evolution. Furthermore, even if the 
inclusive $\mathrm{d}n_{\mathrm{chg}}/\mathrm{d}y$ distribution 
agrees well, this does not mean that more differential distributions
fare as well. As one example, the charged multiplicity distribution
is somewhat broader in \cas.   

As a final technical note, two \pyt instances are used inside \cas.
The main one performs a single $hN$ collision, including decays of so
short-lived particles that they would not have time to interact with
the medium. The other is used to combine this output into an event
record for the whole \hA\ collision, or for a decay. While
taken to be stable in collider physics, the $\mu^{\pm}$, $\pi^{\pm}$,
$K^{\pm}$ and $K_{\mathrm{L}}^0$ are here considered unstable.  

\section{Atmospheric cascade studies}
 
The intention is to develop and provide models that can be used for
detailed studies in a realistic context. Here we will do some simpler first
studies to highlight key features of the new \ang code, and contrast
them with the \cas code, which also has been updated. The main themes
of this section is to compare basic event properties at a fixed energy
or for the whole cascade evolution, and to compare incident proton and
iron either for the same total energy or for the same energy per nucleon.

\subsection{Cross sections}

Key to the evolution of an atmospheric cascade is the \hA\ inelastic
cross section. This definition is not unique, as already mentioned.
Specifically, \ang splits elastic events into true elastic and 
quasi-elastic, depending on whether the nucleus remains intact or
split up, while \cas does not keep track of whether a hadron-nucleon
elastic collision kicks the nucleon out of the nucleus or not. 
What comes out is that the Glauber formalism in \ang gives a larger
elastic cross section than the simpler handling in \cas, and thereby
a larger total one, which has led to some confusion.

\begin{figure*}
\includegraphics[width=0.5\textwidth]{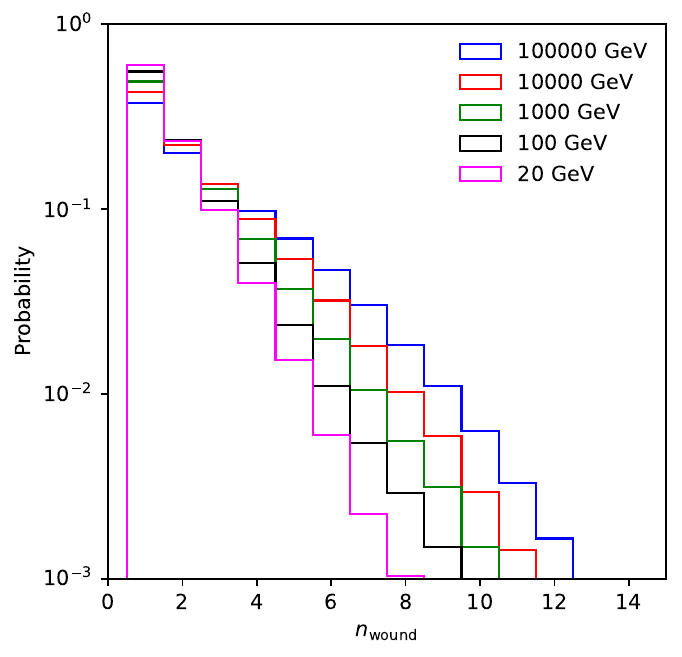}%
\includegraphics[width=0.5\textwidth]{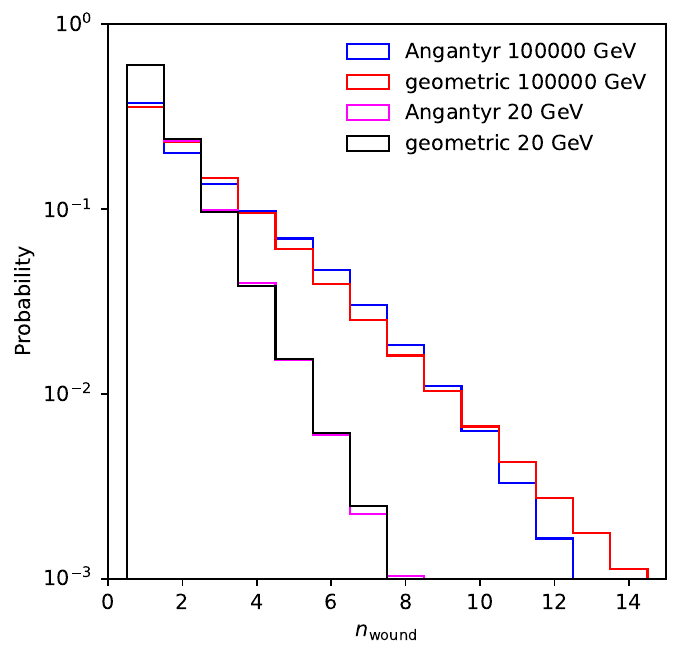}\\[-2mm]
\hspace*{0.23\textwidth}\textit{(a)}\hspace*{0.48\textwidth}\textit{(b)}\\[2mm]
\includegraphics[width=0.64\textwidth]{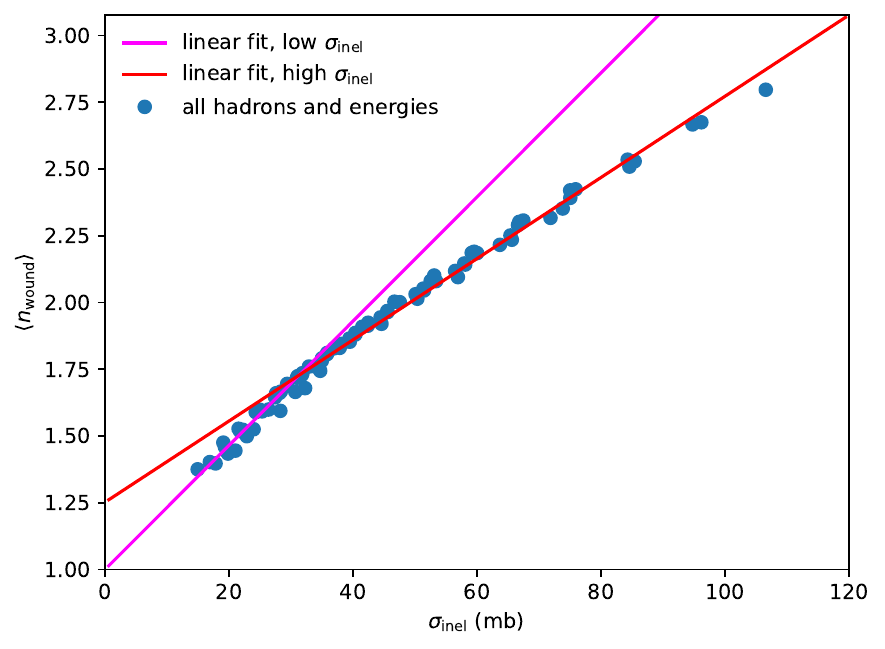}%
\includegraphics[width=0.36\textwidth]{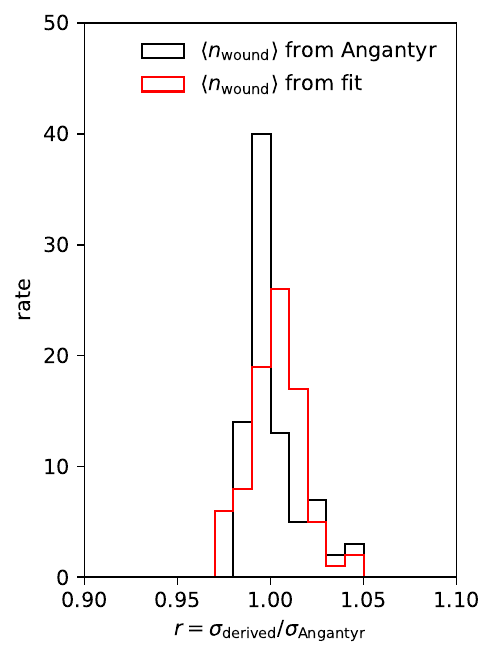}\\[-2mm]
\hspace*{0.28\textwidth}\textit{(c)}\hspace*{0.49\textwidth}\textit{(d)}
\caption{
\textit{(a)} Number $n_{\mathrm{wound}}$ of inelastically wounded
target nucleons in $p$\N14\ at a few $pN$ subcollision energies, 
using full geometry simulation in \ang. 
\textit{(b)} Comparison of $n_{\mathrm{wound}}$ distribution in \ang with
the simpler geometric series in \cas for two energies.
\textit{(c)} The $\langle n_{\mathrm{wound}} \rangle$ value for 84 different
$\sigma_{\mathrm{inel}}$ ones, as described in the text, and the two
resulting linear fit. 
\textit{(d)} Ratio of derived $\sigma_{\mathrm{inel}}^{\hA}$ values to the
\ang ones, for the 84 cases above.}
\label{fig:nWound}
\end{figure*}

In the following we will only consider the inelastic cross section.
Here \cas and \ang have the same input \hN\ cross section, but that
does not guarantee the same subdivision of it. Leaving that aside for
now, the key observable is the number of subcollisions or, equivalently,
the number $n_{\mathrm{wound}}$ of (inelastically) wounded target
nucleons in an \hN\ collision. In Fig.~\ref{fig:nWound}\textit{a} the 
\ang $n_{\mathrm{wound}}$ distribution is shown at a few (per-nucleon) 
collision energies for $p$\N14. This is the starting point for the 
simplified approach in \cas, where the results of the full nuclear 
geometry simulation is replaced by a simple geometric series with the same
$\langle n_{\mathrm{wound}} \rangle$,  Fig.~\ref{fig:nWound}\textit{b}.
While agreement is not perfect, it is unlikely that the approximation
leads to any detectable differences. 

The switching between beams and energies in the atmospheric cascade
evolution requires a convenient para\-metrization of expected
$\langle n_{\mathrm{wound}} \rangle$. To a good first approximation
the beams and energies should not enter individually, but only via
the relevant $\sigma_{\mathrm{inel},\hN}$ value of the two. To this end,
$p$, $\pi^+$, $K^0$, $\phi^0$, $\Sigma^0$, $\Xi^-$ and $\Omega^-$ are
chosen to represent meson and baryon beams of varying strangeness content,
at \hN\ CM frame collision energies $10^{(4 + i) / 3}$, $i$ between
0 and 11, or energies between 21.5~GeV and $10^5$~GeV. The normal
\ang handling of nuclear geometry is only intended to work above 20~GeV,
hence the starting scale. This gives $7 \times 12 = 84$
$\langle n_{\mathrm{wound}} \rangle(\sigma_{\mathrm{inel}})$ values,
Fig.~\ref{fig:nWound}\textit{c}, to be fitted to an interpolating
formula. Inspired by the data, linear least-squares fits are made to
the low- and high-$\sigma_{\mathrm{inel}}$ values separately, where the
former goes up to 30 mb and the latter starts at 25 mb. The former is
also constrained by $\langle n_{\mathrm{wound}} \rangle \to 1$ for
$\sigma_{\mathrm{inel}} \to 0$. Henceforth
$\langle n_{\mathrm{wound}} \rangle(\sigma_{\mathrm{inel}})$ is chosen as
the smaller value of the two linear fits.

In Fig.~\ref{fig:nWound}\textit{d} eq.~(\ref{eq:sigmaInelhA}) is
tested by comparing $\sigma_{\mathrm{inel}}^{h\N14}$ ratios. The
numerators are the ones derived from eq.~(\ref{eq:sigmaInelhA}) using
the $\langle n_{\mathrm{wound}} \rangle$ values in \ang and
in the fit, respectively, while the denominator is the $h\N14$
cross section value provided directly by \ang. As can be seen,
the spread for the 84 data points is not appreciably worse from the
fit than for \ang itself, which means that the fit ought to be a good
predictor for other beams and energies.

Similar parametrizations as for \N14\ above has been done for
14 other common nuclear targets, from $^2$H to $^{208}$Pb, including
\O16\ and \Ar40.  (For targets from $^{84}Kr$ onwards the highest
energy point is omitted.) Interpolation is performed for intermediate 
nuclei. 

\begin{figure}
\includegraphics[width=\columnwidth]{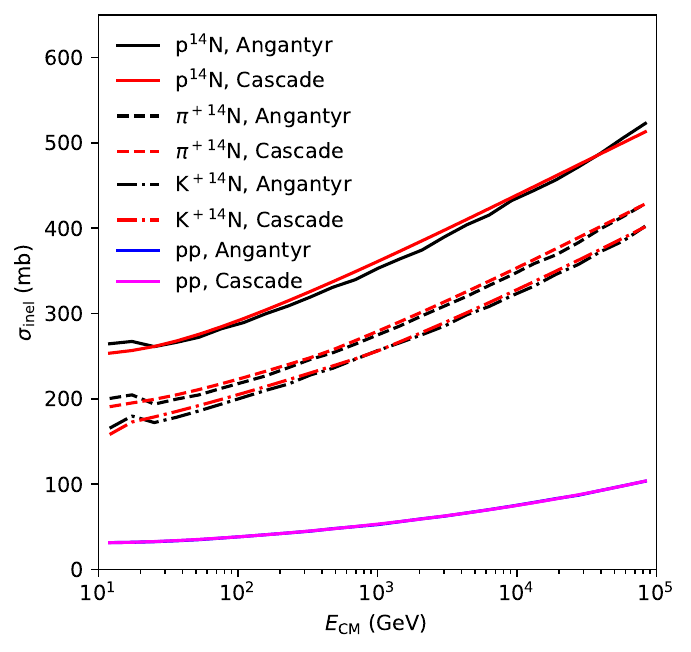}
\caption{Energy dependence of the inelastic cross section for some
common particles hitting \N14, comparing \ang and \cas.
\pp\ collisions are included as a reference.}
\label{fig:sigmaInel}
\end{figure}

The $\sigma_{\mathrm{inel}}$ end result is illustrated in
Fig.~\ref{fig:sigmaInel}. The Angantyr values have fluctuations
both from Monte Carlo statistics and from the internal geometry
algorithms, while the \cas curves are based on the fitted
$\langle n_{\mathrm{wound}} \rangle$, so are smooth from the onset.
Nevertheless overall good agreement is found. 

The \ang jump at
20~GeV is caused by two different geometry handlings being used
at low and at high energies. At high ones, the full Glauber formalism
is used, with fluctuating nucleon sizes. The tuning of free parameters
in such an approach becomes unstable at low energies, however.
Instead a simpler framework is used, where the nucleons have fixed sizes.
This gives of the order of 5\% higher cross sections below 20~GeV than
\cas does. 
 
One may note that the detailed event-by-event setup of nuclear geometry,
as opposed to the simple geometric series, makes \ang slower than \cas.
In the region just above 20~GeV this may give as much as a factor of
three slowdown. At higher energies the other generation steps take
more time, so proportionately speaking the nuclear geometry becomes
less important, even if it does not decrease in absolute terms.
To improve the situation, a set of nuclear configurations may be stored
and reused, with different incoming beams. Below 20 GeV a simplified
geometry handling is used in \ang, which reduces the \ang-to-\cas
discrepancy.

Agreeing inelastic cross sections does not necessarily mean agreeing
partial cross sections, not even for the $pp$ starting point.
For instance, at 1~TeV the SaS ansatz for inelastic processes is
36.33 mb nondiffractive, two times 5.40 mb single diffractive, and
5.94 for double diffractive, giving a total of 53.07 mb. The \ang
Glauber handling instead gives 36.78 mb, twice 5.63 mb, and 5.01 mb,
for a total of 53.04. That is, Glauber gives more single diffraction
and less double one in this case.

\subsection{Event properties}

\begin{figure*}
\includegraphics[width=0.5\textwidth]{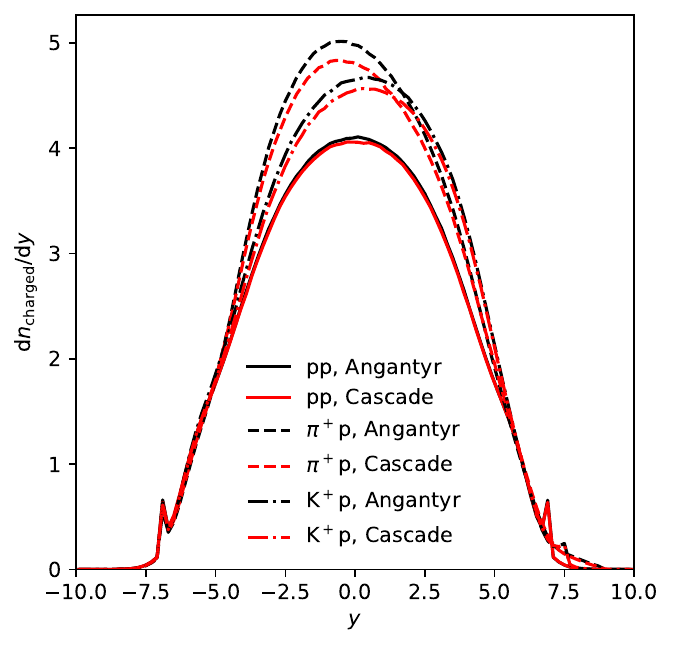}%
\includegraphics[width=0.5\textwidth]{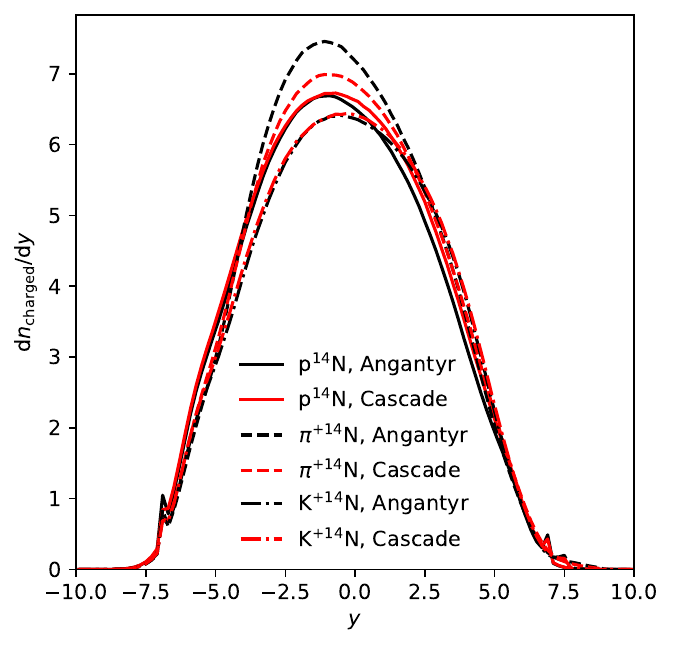}\\[-2mm]
\hspace*{0.23\textwidth}\textit{(a)}\hspace*{0.48\textwidth}\textit{(b)}
\caption{Rapidity distribution of charged particles, for a 1000 GeV
hadron--nucleon CM energy, in the hadron--nucleon rest frame. 
\textit{(a)} Proton target.
\textit{(b)} Nitrogen target.}
\label{fig:dndyRef}
\end{figure*}

Event shapes are important for the cascade evolution. In
Fig.~\ref{fig:dndyRef}\textit{(a)} the CM-frame charged-particle rapidity
spectrum for inelastic events is compared between $p$, $\pi^+$ and
$K^+$ beams on a $p$ target. The event activity is directly related to 
the average number of multiparton interactions, which is given by the 
ratio of the MPI cross section to the inelastic cross section,
\begin{equation}
\langle n_{\mathrm{MPI}} \rangle = 
\frac{\sigma_{\mathrm{MPI}}}{\sigma_{\mathrm{inel}}} ~.
\end{equation} 
For the sake of simplicity we here avoid discussing the split between 
diffractive and nondiffractive  topologies \cite{Rasmussen:2015qgr}.
Another potential complication is that the MPI cross section is
divergent in the $p_{\perp} \to 0$ limit, which is addressed by
introducing a $p_{\perp 0}$ regularization parameter. Its value need
not be the same for all particle combinations, but is for simplicity
assumed to be so.

The $\sigma_{\mathrm{MPI}}$ numerator involves the integration of 
the perturbative parton--parton interaction rate with the parton 
distributions of the two incoming hadrons. By definition the 
$x$-weighted flavour-summed integral is unity for all hadrons, 
but the shape may differ. One may especially note that the $\pi^+p$ 
spectrum peaks at $y < 0$, reflecting differences in the $x$ 
distributions of $\pi$ and $p$. Here it is relevant to recall that
only $p/n$ PDFs are well known, with some limited data on $\pi^{\pm}$
and $K^{\pm}$, and nothing at all on all the other hadrons.

The $\sigma_{\mathrm{inel}}$ denominator is lower for $\pi^+p$ than 
for $pp$, which translates into a higher multiplicity for the former. 
$K^+p$ has an even lower $\sigma_{\mathrm{inel}}$ than $\pi^+p$, but 
the $\overline{s}$ quark in the kaon is expected to take a larger 
fraction of the total momentum than the quarks in the pion, leaving 
less to low-$x$ gluons, thus giving a reduced $\sigma_{\mathrm{MPI}}$
and $\langle n_{\mathrm{MPI}} \rangle$. The $K^+p$ process therefore
lies below $\pi^+p$, but still above $pp$.

Zooming in on the $pp$ curves for a moment, the tiny difference between
\ang and \cas comes from a combination of two effects. One is a small
divergence in the nondiffractive event generation. The other is the
single and double diffractive cross section mix, while the distributions
per diffractive event are the same.

The picture is changed when switching to a \N14\ target,
Fig.~\ref{fig:dndyRef}\textit{(b)}. One visible effect is that the
distribution maxima are shifted appreciably towards negative
rapidities, i.e.\ in the target direction. This is natural
consequence of having 14 nucleons in the target, each with
$p_z = -500$ GeV, while there is only one hadron with $p_z = +500$ GeV.
So the more nucleons are hit, the more activity in the target
direction. Technically the way \ang and \cas achieve this are
different, but the outcome similar. 

The possibility of having more than one wounded nucleon in
$h$\N14\ collisions also increases the charged multiplicity
relative to an $hp$ one. And since
$\sigma_{\mathrm{inel}}^{pp} > \sigma_{\mathrm{inel}}^{\pi^+p}
> \sigma_{\mathrm{inel}}^{K^+p}$ it follows that
$\langle n_{\mathrm{wound}}^{p\N14} \rangle
> \langle n_{\mathrm{wound}}^{\pi^+\,\N14} \rangle
> \langle n_{\mathrm{wound}}^{K^+\,\N14} \rangle$. 
This allows the $p$\N14\ multiplicity to overtake the 
$K^+\,\N14$ one, and close in on the $\pi^+\,\N14$ one. 

\begin{figure*}
\includegraphics[width=0.5\textwidth]{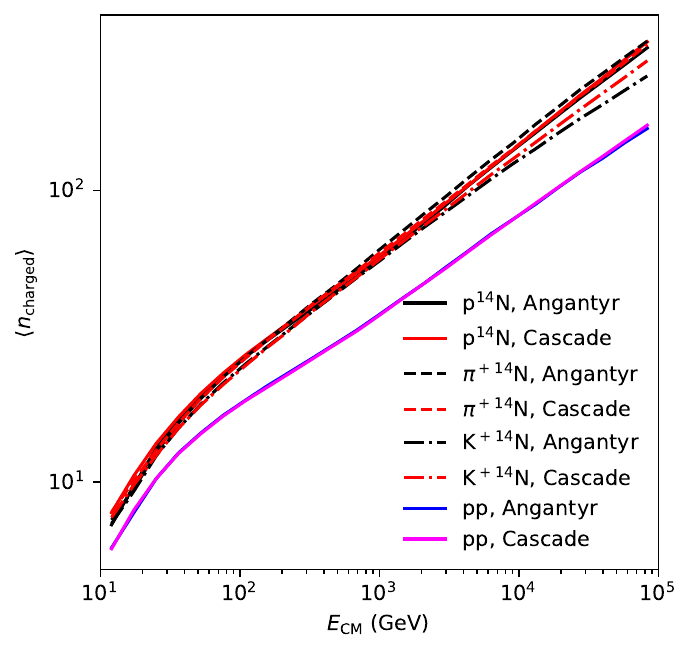}%
\includegraphics[width=0.5\textwidth]{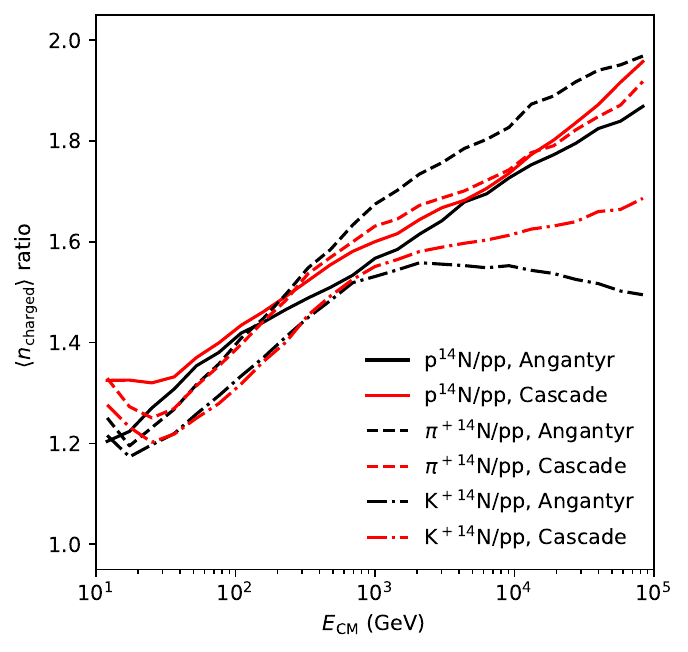}\\[-2mm]
\hspace*{0.23\textwidth}\textit{(a)}\hspace*{0.48\textwidth}\textit{(b)}
\caption{\textit{(a)} Hadron--nucleon CM energy dependence of the average
charged multiplicity for $p/\pi^+/K^+$ beams on a \N14\ target, comparing 
\ang and \cas. \pp\ collisions are included as a reference.
\textit{(b)} Ratio of the $h\N14$ multiplicities to the \pp\
\ang + \cas average one.}
\label{fig:nchevol}
\end{figure*}

The collision energy dependence of the charged multiplicity is shown
in Fig.~\ref{fig:nchevol}\textit{(a)}. For $pp$ \ang and \cas overlap
closely --- a good sanity check --- while the six $h\N14$ curves also
are clustered. To improve the resolution,
Fig.~\ref{fig:nchevol}\textit{(b)} shows the $h\N14$ curves normalized
to the average $pp$ ones. The general trend of an increasing ratio
is a natural consequence of an increasing number of subcollisions,
which in its turn comes from increasing cross sections. The hierarchy
with $\pi^+\,\N14$ being above and $K^+\,\N14$ below $p\,\N14$ is also
recognized from Fig.~\ref{fig:dndyRef}\textit{(a)}. Notable is that
$\pi^+\,\N14$ and $K^+\,\N14$ are closer in \cas than in \ang,
which likely is a consequence of the flipping of interacting hadron
in the \cas nuclear collision chain. Here we thus encounter a clear
manifestation of the simplified \cas modelling, even if it may be
largely masked by the mix of collision processes in an atmospheric
cascade. Another weird feature is the larger ratio at the lowest 
energies, especially for \cas, which has no obvious explanation. 
This is in a transition region, from a simpler longitudinal-string 
description at low energies to the MPI-based one at higher energies, 
so the cause may be how this affects secondary collisions. Anyway, 
on the absolute scale of Fig.~\ref{fig:nchevol}\textit{(a)}, 
it does not look so bad.

\begin{figure}
\includegraphics[width=\columnwidth]{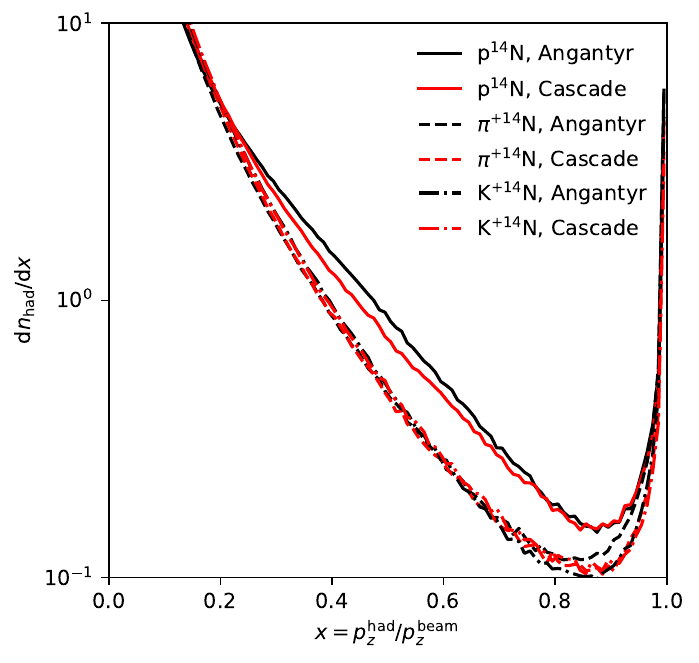}
\caption{Longitudinal momentum fraction of stable hadrons for 
$p/\pi^+/K^+$ beams on an \N14\ fixed target, for a 1000 GeV 
hadron--nucleon CM energy, comparing \ang and \cas.}
\label{fig:xFrac}
\end{figure}

While the overall charged multiplicity in collisions is relevant, 
the evolution of an atmospheric cascade is largely driven by the 
most high-energy hadrons, whether charged or not. Here 
Fig.~\ref{fig:xFrac} shows three interesting features. Firstly,
all curves have a peak near $x = 1$, coming from single diffraction,
with the diffracted system in the target region and the beam particle
scattering quasi-elastically. 
Secondly, the proton beam gives more hadrons at intermediate $x$ 
than the $\pi^+/K^+$ ones, which is a direct consequence of the \pyt
beam remnant handling, where a diquark is allowed to take a larger
momentum fraction than a single quark. This is further augmented by a 
harder fragmentation function into baryons than into mesons,
from mass effects. (Potentially also from dynamical mechanisms
\cite{Fieg:2023kld}.) And thirdly that \ang tends to give more
medium- and high-$x$ hadrons than \cas, even if the differences
are not dramatic.

The comparison plots do not go below 10~GeV, but nothing dramatically
unexpected happens for lower energies. Cross sections are rather flat, since
the decreasing reggeon term and the increasing pomeron one are of
comparable magnitude. As a consequence, the number of wounded
nucleons in a \hA\ collision would remain fairly flat. But there
is a catch, that this number is kinematically restricted by the 
energy at disposal for inelastic hadron--nucleon subcollisions. 
This requirement hits harder in the \ang modelling than in the 
\cas one, and therefore \cas generates a somewhat higher
multiplicity at low energies. This difference tends to be around
or above 10\% per collision. Even given the $\sim$5\% higher \ang
inelastic cross section, overall multiplicities are higher in \cas.
The difference is most notable in the number of low-energy nucleons,
and have little impact on higher-energy muons or neutrinos.
In the future the modelling in the low-energy region should be 
contrasted with fixed-target data, and be adjusted as called for.

In passing we note that \cas is slower than \ang  in the handling 
of the lower-energy stages of the cascade, by about 25\%, and this
is where most of the total time is spent. 

\begin{figure*}
\includegraphics[width=0.5\textwidth]{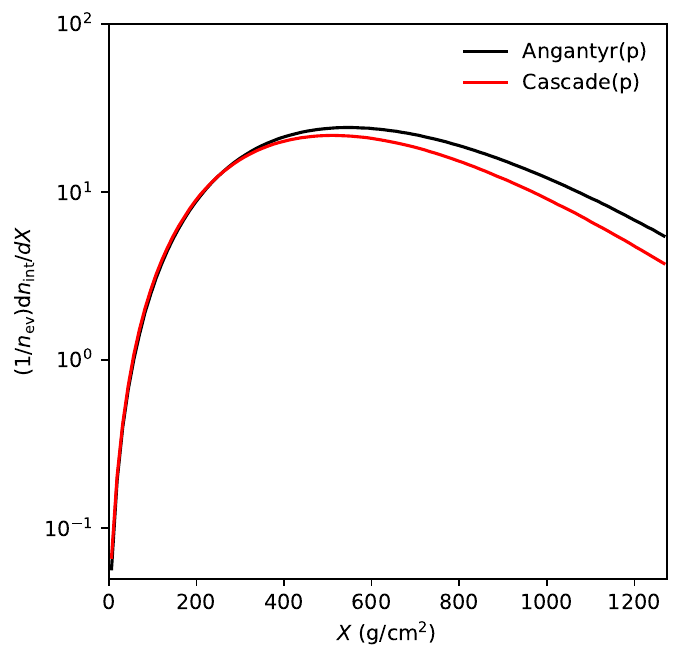}%
\includegraphics[width=0.5\textwidth]{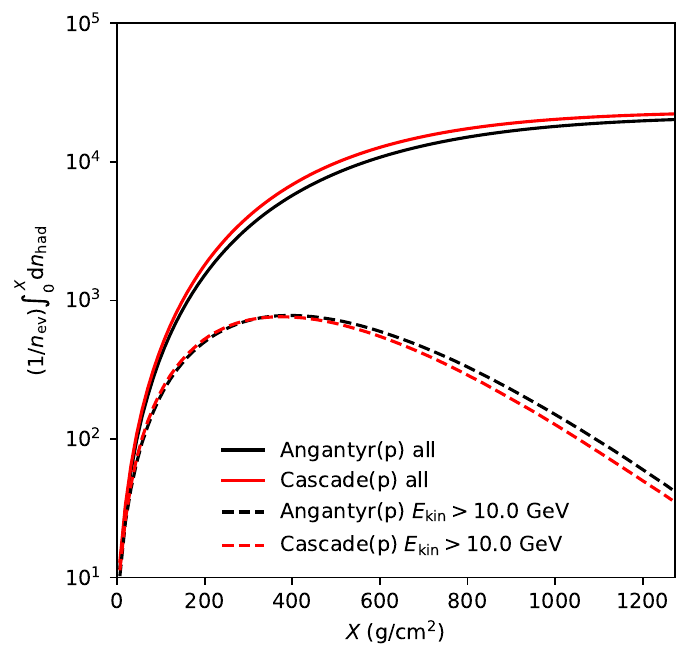}\\[-2mm]
\hspace*{0.23\textwidth}\textit{(a)}\hspace*{0.48\textwidth}\textit{(b)}\\[2mm]
\includegraphics[width=0.5\textwidth]{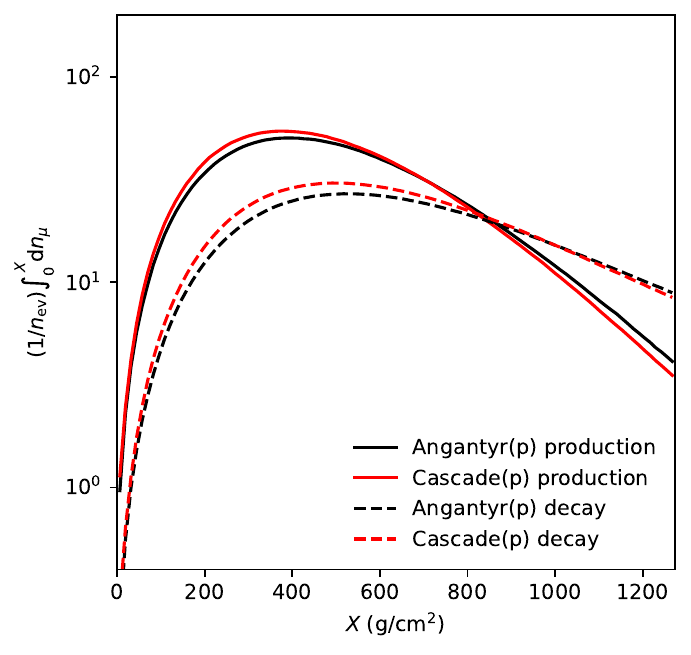}%
\includegraphics[width=0.5\textwidth]{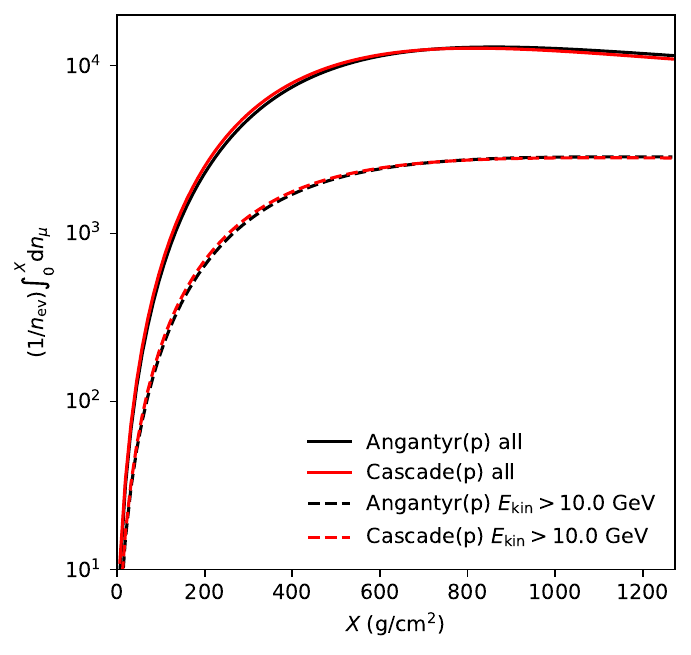}\\[-2mm]
\hspace*{0.23\textwidth}\textit{(c)}\hspace*{0.48\textwidth}\textit{(d)}
\caption{
Example of properties of a hadronic cascade induced by a vertically
incoming proton of $10^6$~GeV initial energy, with a lower cutoff of
interactions at 0.5~GeV kinetic energy.
\textit{(a)} Atmospheric depth $X$ of nuclear interactions.
\textit{(b)} Number of hadrons at a depth $X$, all and those with a
kinetic energy above 10 GeV.
\textit{(c)} Muon production and decay rates at a depth $X$.
\textit{(d)} Number of muons at a depth $X$, all and those with a
kinetic energy above 10~GeV.}
\label{fig:fullcasc}
\end{figure*}

\subsection{Complete cascades}

A simplified model of the atmosphere is used to study the hadronic 
part of cosmic ray cascades, as a freestanding test of the code 
prior to inclusion in more sophisticated frameworks. This model 
is based on the same approach as outlined in
\cite{Sjostrand:2021dal}. In it, the atmospheric density  follows 
an exponential distribution as a function of the height $h$ above 
the (sea-level) surface, $\rho(h) = \rho_0 \, \exp(- h/ H)$, 
where $\rho_0 = 1.225$ kg/m$^3$ and $H = 10.4$ km. Atmospheric
cascades are traced from a 100 km height, neglecting interactions
above that. The composition is chosen to be 78.5\% nitrogen, 
21\% oxygen and 0.5\% argon, by number density of nuclei. (Often 
the composition is quoted in terms of number of molecules, and then 
argon is 1\% since it is a noble gas while nitrogen and oxygen form 
molecules N$_2$ and O$_2$.)

\begin{figure*}
\includegraphics[width=0.5\textwidth]{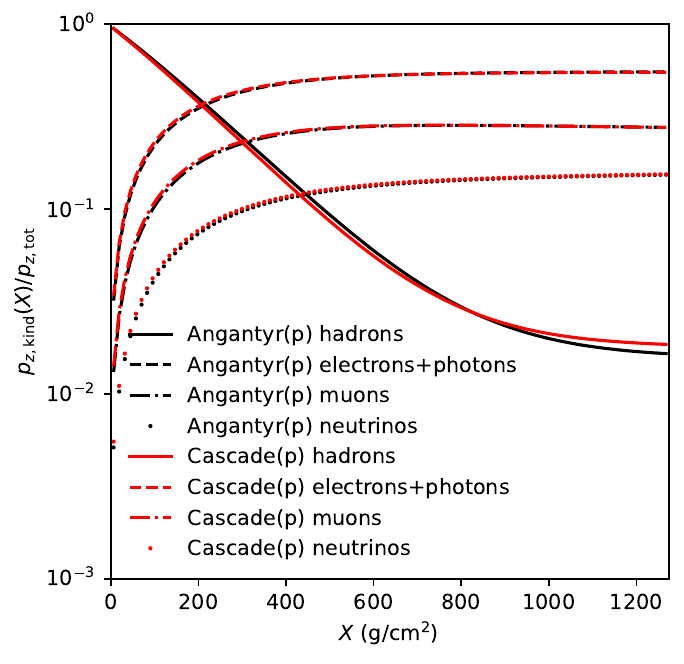}%
\includegraphics[width=0.5\textwidth]{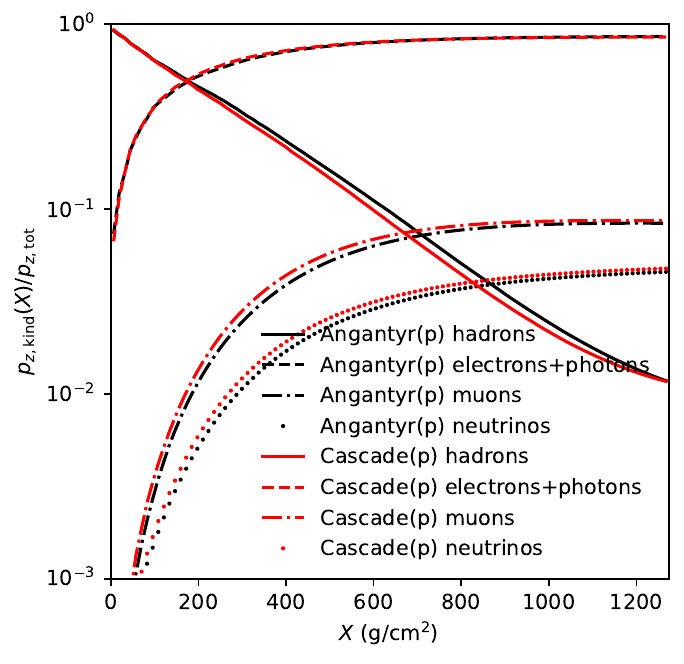}\\[-2mm]
\hspace*{0.23\textwidth}\textit{(a)}\hspace*{0.48\textwidth}\textit{(b)}
\caption{Fraction of the initial momentum carried by hadrons,
by electrons and photons, by muons, and by neutrinos, as a function
of $X$. The initial proton has an energy of \textit{(a)} $10^4$~GeV
and  \textit{(b)} $10^8$~GeV, respectively.}
\label{fig:energyshare}
\end{figure*}

There are also some important simplifications. Electromagnetic
cascades are not simulated so, once an $e^{\pm}$ or a photon is
produced, nothing more happens to it. Also muon interactions are
absent, but muons can decay. The curvature of the earth is neglected, 
which is only a problem for near-horizon cascades. Bending of charged
particles by the earth's magnetic field is also neglected.  

Example of properties in such an atmospheric cascade are shown in
Fig.~\ref{fig:fullcasc}, for a $10^6$~GeV incoming proton. Overall
\ang and \cas give similar properties, especially at the beginning
of the cascade. At later stages the two diverge, with \cas showing
a smaller interaction rate with the atmosphere than \ang. This runs
somewhat contrary to \cas having a larger particle production rate,
even in the early region where interaction rates agree. But we have
already noted that low-energy interactions produce approximately 
10\% more particles in \cas than in \ang. This is visible e.g. in 
the overall number of hadrons, Fig.~\ref{fig:fullcasc}\textit{(b)}. 
More particles translates into less energy per particle, which is 
visible in the rate of hadrons above 10~GeV. Muons are a bit of a 
special case, since larger production also means more decays, 
Fig.~\ref{fig:fullcasc}\textit{(c)}. Lower energy also means less 
time dilation and thus higher probability to decay before reaching 
the surface, thereby giving a harder spectrum for those that do.
The competing effects work out so that muon rates agree well between
\ang and \cas, Fig.~\ref{fig:fullcasc}\textit{(d)}, both for all and 
for higher-energy muons.

Fig.~\ref{fig:energyshare} shows how the initially hadronic
energy leaks over to $e^{\pm}/\gamma$, to muons, or to neutrinos.
The production and decay of $\pi^0$ is the main reason why the
electromagnetic part soon dominates. The larger early rate of hadrons 
in \cas also translates into a larger early loss of hadronic energy. 
Later \ang catches up or even overtakes, but overall differences
are tiny.

Obviously results depend on the energy of the cascade initiator. Since 
most cascades eventually are dominated by low-energy particles, the
generated multiplicity grows almost linearly with the initial energy:
for the factor 10000 step between $10^4$ and $10^8$~GeV the
multiplicity increases by almost a factor 5000.
A changed energy also affects the balance between collisions and
decays. Notably higher-energy $\pi^{\pm}$ and $K^{\pm}$ get increased 
chances to interact before they decay, owing to time dilation. This 
leads to a larger fraction of the initial energy going to the 
electromagnetic component, and lesser to muons and neutrinos, 
Fig.~\ref{fig:energyshare}. 

\subsection{Nuclear beams}

The area where \ang excels over \cas is that it can handle \AA\
collisions, e.g.\ from incoming iron nuclei, which usually is chosen
as a contrast to the incoming proton case studied so far. There are 
two specific applications of interest.

\begin{figure*}
\includegraphics[width=0.5\textwidth]{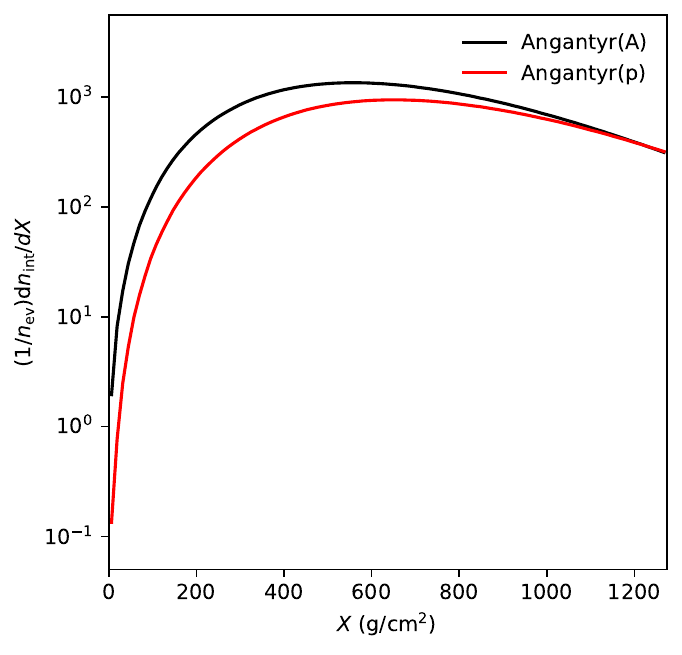}%
\includegraphics[width=0.5\textwidth]{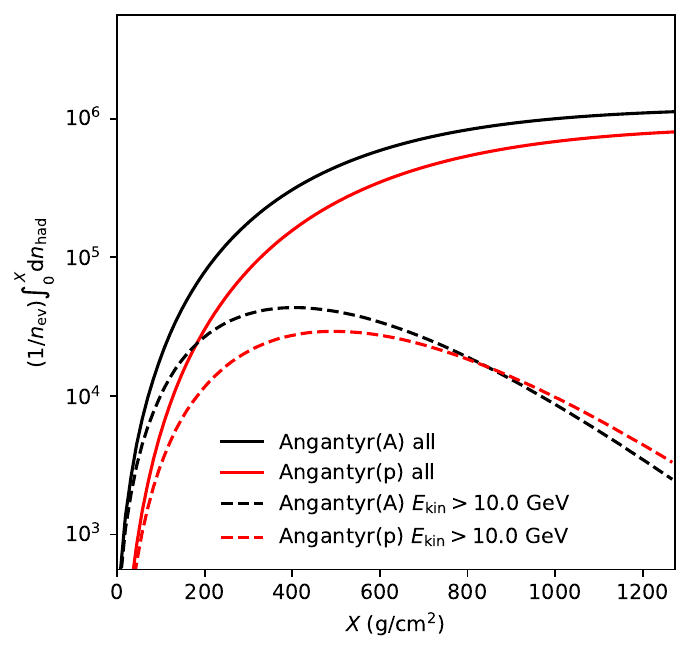}\\[-2mm]
\hspace*{0.23\textwidth}\textit{(a)}\hspace*{0.48\textwidth}\textit{(b)}\\[2mm]
\includegraphics[width=0.5\textwidth]{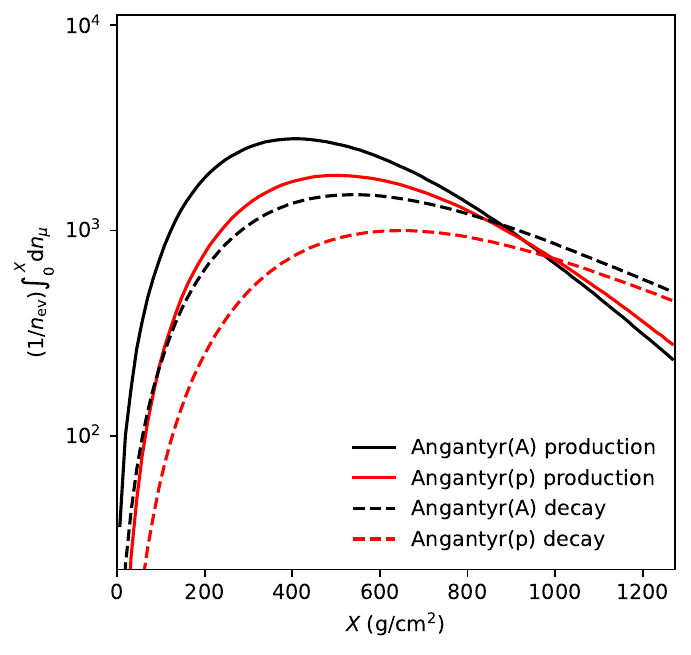}%
\includegraphics[width=0.5\textwidth]{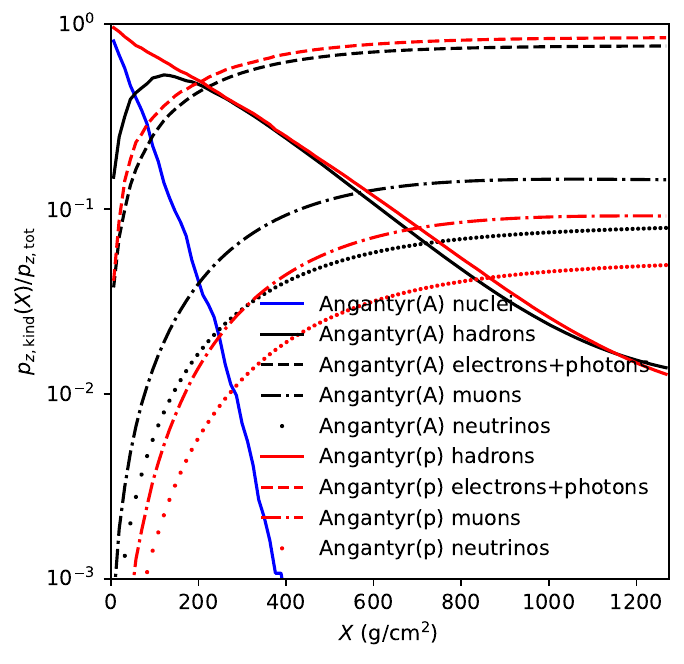}\\[-2mm]
\hspace*{0.23\textwidth}\textit{(c)}\hspace*{0.48\textwidth}\textit{(d)}
\caption{
Example of properties of a hadronic cascade induced by a vertically
incoming iron nucleus (``Angantyr(A)'') with an energy of $10^6$~GeV
per nucleon or a single proton (``Angantyr(p)'') with an energy of
$56 \cdot 10^6$~GeV.
\textit{(a)} Atmospheric depth $X$ of nuclear interactions.
\textit{(b)} Number of hadrons at a depth $X$, all and those with a
kinetic energy above 10 GeV.
\textit{(c)} Muon production and decay rates at a depth $X$, all and 
those with a kinetic energy above 10~GeV.
\textit{(d)} Fraction of the initial momentum carried by nuclei (only
applicable for the iron beam), by hadrons, by electrons and photons,
by muons, and by neutrinos, as a function of $X$.}
\label{fig:ironVSp}
\end{figure*}

Firstly, from the measured energy of the electromagntic cascade it is
possible to estimate the original energy of the incoming projectile.
A comparison of the cascade evolution between $p$ and \Fe56\
projectiles of the same energy allows to understand in what way they
differ, and thereby to gain insight on the composition of high-energy
cosmic rays. Such a comparison is shown in Fig.~\ref{fig:ironVSp}.
The evolution of the iron nucleus cascade starts earlier than that of
the $p$, owing to its larger cross section, and these differences
persist. Thus more hadrons, electron/photons, muons and neutrinos
are produced in the iron cascade. This decreases the average energy
per particle, and as we see this eventually means more hadrons above
10~GeV for the proton cascade. For muons and neutrinos, however,
the iron cascade is on top also there. This is related to their
larger fraction of the total momentum, mainly at the expense of the
electromagnetic cascade part.

\begin{figure*}
\includegraphics[width=0.5\textwidth]{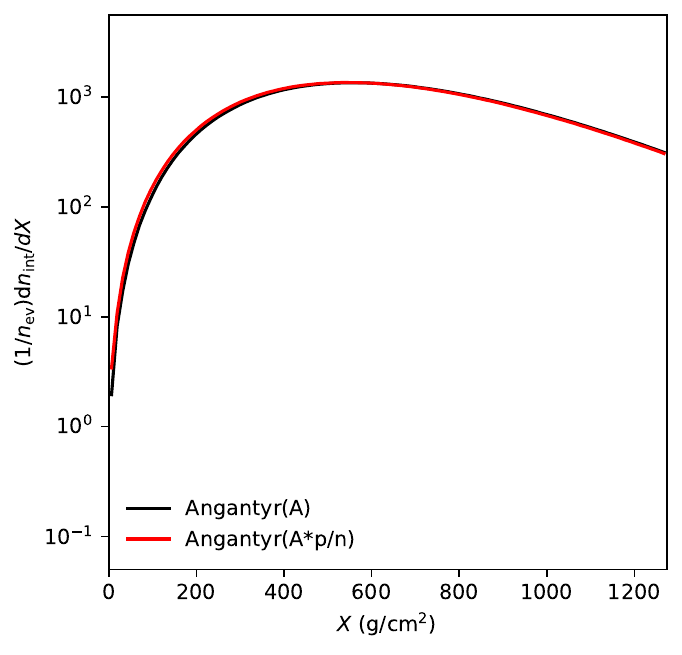}%
\includegraphics[width=0.5\textwidth]{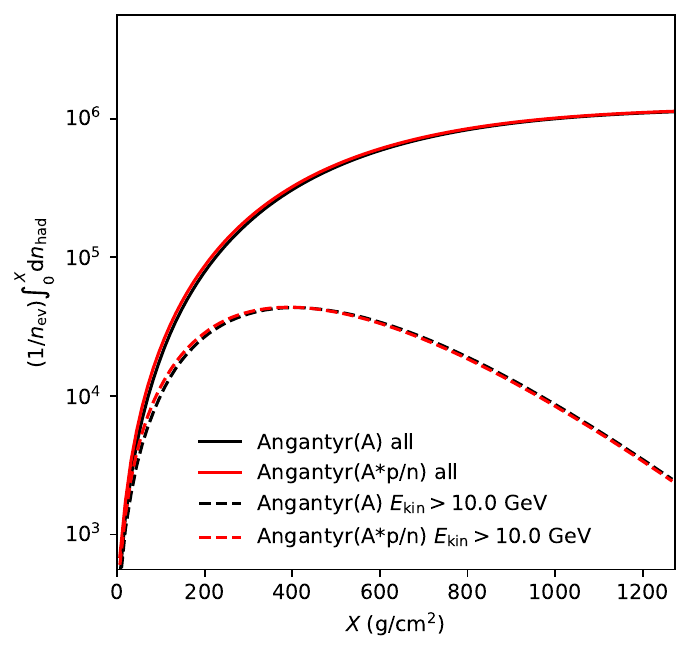}\\[-2mm]
\hspace*{0.23\textwidth}\textit{(a)}\hspace*{0.48\textwidth}\textit{(b)}\\[2mm]
\includegraphics[width=0.5\textwidth]{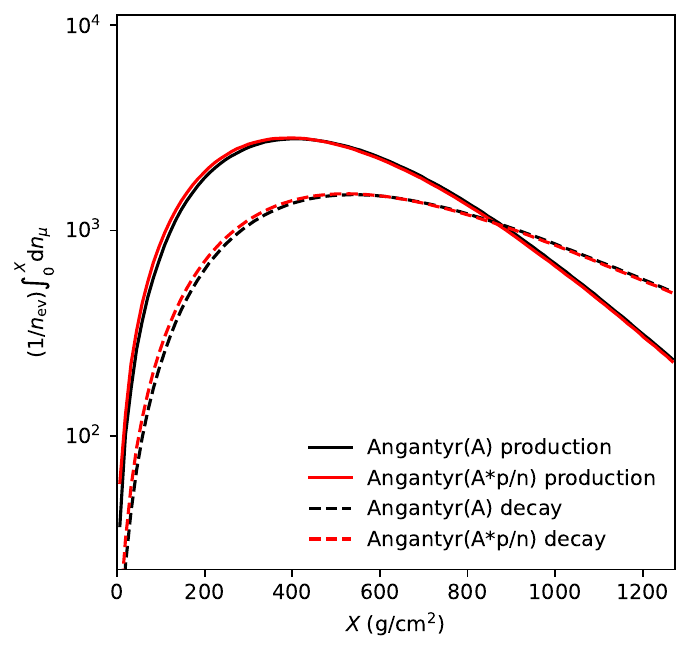}%
\includegraphics[width=0.5\textwidth]{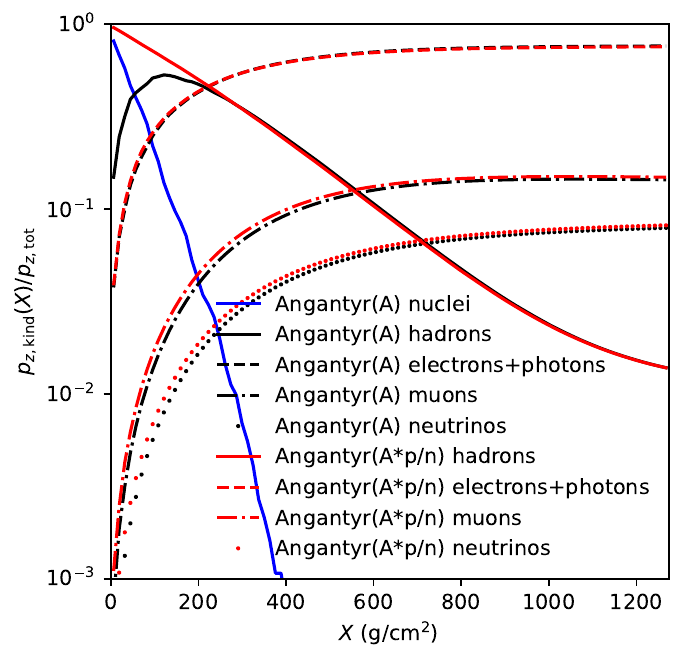}\\[-2mm]
\hspace*{0.23\textwidth}\textit{(c)}\hspace*{0.48\textwidth}\textit{(d)}
\caption{
Example of properties of a hadronic cascade induced by a vertically
incoming iron nucleus (``Angantyr(A)'') or 56 nucleons
(``Angantyr(A*p/n)''), with an energy of $10^6$~GeV per nucleon.
Frame captions as in Fig~\ref{fig:ironVSp}.}
\label{fig:ironVSindep}
\end{figure*}

Secondly, a standard trick for codes that cannot handle a incoming 
nucleus is to model it as a superposition of independent nucleon 
cascades, where the nucleons share the original nuclear energy evenly
\cite{Engel:1992vf}. This is used e.g.\ in \textsc{Sibyll}, and would 
offer a possibility for \cas as well. Fig.~\ref{fig:ironVSindep}
presents precisely such a comparison where an incident single iron
nucleus is compared with 56 separate nucleons, 26 $p$ and 30 $n$,
with a $10^6$~GeV energy per nucleon in both cases. As can be seen,
the similarity of event properties is striking; event better than
one might have guessed. The larger cross section per iron nucleus,
times the larger number of wounded nucleons per collision, is expected
to compensate fairly well for only being one rather than 56 separate
ones, cf. eqs.~(\ref{eq:sigmaTothA}) and (\ref{eq:sigmaInelhA}).
But, in the \ang model (and by implication also the \cas one), wounded
nucleons beyond the first one contribute less additional multiplicity
than an independent nucluon does. One therefore would expect less
particle production in the iron case. Indeed there are some small
such effects, notably in the reduced early production of muons, 
as the nucleus and hadrons retain a somewhat larger fraction of the
full energy. But, by the time the cascade comes to the lower
atmospheric layers, differences are gone. 

\begin{figure}
\includegraphics[width=\columnwidth]{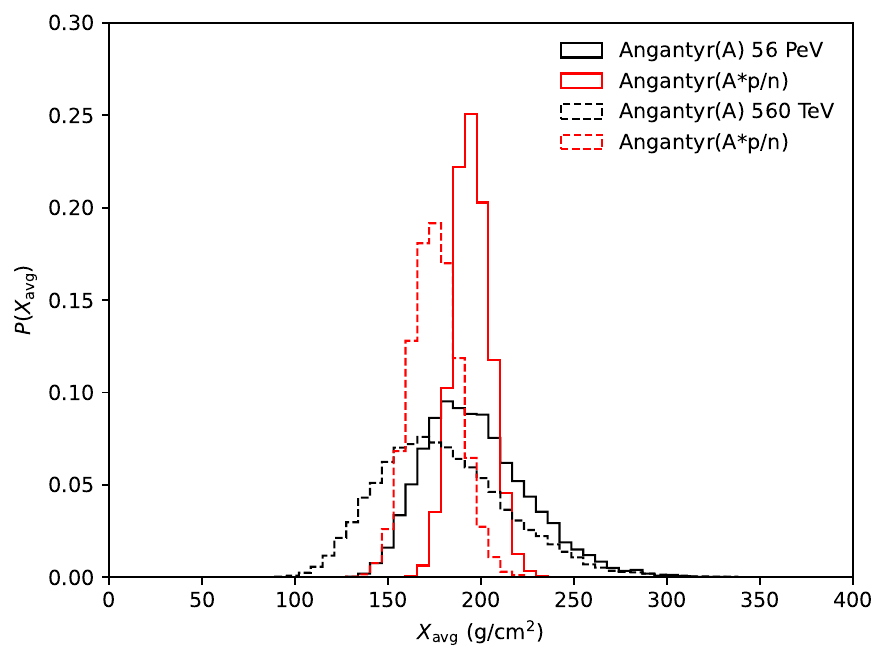}
\caption{The distribution of the energy-weighted average production
  depth of electrons and photons in the hadronic cascade,
  for a vertically incoming iron
  nucleus (``Angantyr(A)'') compared to 56 nucleons
  (``Angantyr(A*p/n)'') for two different energies, $56\times10^6$~GeV
  (solid lines) and $56\times10^4$~GeV (dashed lines).}
\label{fig:xAvg}
\end{figure}

One small comment is that \ang, in each collision, puts all non-wounded
nucleons into a new remnant nucleus. This then does not take into account
that the remnant is likely to be left in an excited state, from which
further nucleons could evaporate. Then \ang would underestimate
the rate at which the original nucleus splits into separate cascades.   
(There may also be evaporation from the target nucleus but, since the 
target is at rest, with so low energies that these nucleons will not 
propagate far.) So, had we noted significant differences above, we could
have attributed part of them to the absence of this mechanism. As it
happens, we do not even have to use this as a partial explanation of
differences.

So far the study confirms the standard trick of handling an incoming
nucleus as separate nucleons, implying that \cas can be used equally
well as \ang for such tasks. There are still differences, however,
which was already noted in \cite{Engel:1992vf}, namely that the
distribution in $X_{\max}$ --- the position of the maximum of the
electromagnetic shower --- is wider in the case of a full nucleus, as
compared to $A$ individual nucleons. This is expected as taking the
average of $A$ individual projectiles naturally will have smaller
fluctuations than a single projectile. Since we do not simulate
electromagnetic cascades in \pyt, we can not easily access $X_{\max}$.
Instead we show in Fig.~\ref{fig:xAvg} the distribution of the
energy-weighted average production depth of electrons and photons
in the hadronic cascade,
$X_{\mathrm{avg}}$. This should be well correlated with $X_{\max}$,
and we find, qualitatively, the same effect. In Fig.~\ref{fig:xAvg} we
show the effect for two different energies, $56\times10^6$~GeV and
$56\times10^4$~GeV. It is interesting to note that in both cases
the distributions are somewhat wider for the lower energy and also
peak a bit earlier. The reason is that pions, which are the
main source of electrons and photons, although produced
earlier due the increase in cross section at higher energies, 
also are more time-dilated and more often interact hadronically again,
before they have time to decay.

\section{Summary and outlook}

The main thrust of this article has been to enable variable beams and 
energies in the \ang framework of hadron--nucleus and nucleus--nucleus
collisions, extending on the \pyt framework for hadron--hadron ones. 
This has required significant development of the code, with  numerous
issues to be solved, mainly administratively, but also in terms of
physics modelling. Much of this modelling has been taken over from
the code extensions introduced to support \cas. That development 
is now finished, and the code will be made public in an upcoming 
release of \pyt.  

Also the \cas code has been updated, even if at a more modest scale.
\cas was originally conceived as a ``poor man's'' alternative to 
\ang, offering the flexibility that \ang had not. It was anchored to 
a few key observables of \ang at fixed energies and beams, notably
total and inelastic cross sections, the number of wounded nucleons, 
and the event activity associated with secondary wounded nucleons. 
In particular the \ang bookkeeping of the former has been misleading
and required adjustments. Now overall good agreement with the updated
\ang has been restored, at least for higher energies. In this sense 
\cas has served its purpose and could be retired, but its more robust
structure makes it a useful backup and cross-check.

A key observation is that \ang simulation of incoming heavy nuclei, 
here represented by iron, does not give a cascade evolution that
differs significantly from what is obtained if each incoming nucleon
is allowed to cascade independently. This implies that \cas, designed
only to handle a single hadron on a nuclear target, is easily extended
also to incoming nuclei. 

A small caveat is that we here only considered collisions in air.
But \pyt is already now used in detector simulation programs like
\textsc{Geant~4}, even if in older incarnations, as mentioned in the
introduction. And one of the objectives of the \ang development work
is to make \ang a valid replacement to those earlier codes, which are
in Fortran rather than C++. Sampling calorimeters can here give a
quite precise reconstruction of the early shower development, which we
have not studied closer in this article, and here the \ang simulation
could have advantages over the \cas one. This is a minor point for
$pp$ collisions at the LHC, where the primary production  of deuterium,
tritium and helium is observed to be at very low levels. For \AA\ ones
the likelihood of nuclear remnants increases, in particular in the
forward direction.

The framework developed here could also have other applications.
Incoming neutrinos, from the cosmos or from man-made accelerators,
can undergo a charged or neutral current interaction inside the
detector, which leads to a subsequent cascade. The primary interaction
can then be handled by the basic ``old'' \pyt machinery, and the
subsequent evolution by the codes developed here. (With or without
the help of \textsc{Geant} or other tracking codes.) Also, a fraction 
of high-energy photons interact hadronically with the target material 
--- photoproduction --- which leads to new hadronic cascades, where 
\pyt could be used both for a primary and for subsequent interactions.

An important aspect to remember, however, is that neither \ang nor
\cas has been tuned to low-energy data. We have noted that this is
a region where the two implementations give somewhat different
predictions. In the near future it will be relevant to compare
with such data, available mainly via \textsc{Rivet}
\cite{Bierlich:2024vqo}, and use these comparisons to improve the
modelling. Notably, collider comparisons tend to concentrate on
particle production at central rapidities, while here the forward
region is more important.

In summary, while the current article represents a big step towards
making \pyt relevant for a wide variety of particle physics collision
processes, we can foresee further developments to further improve
the scope and accuracy of the frameworks presented here.

\begin{acknowledgement}
Work supported in part by the Swedish Research Council, contract numbers
2016-05996 and 2020-04869.
\end{acknowledgement}

% BibTeX users please use
\bibliographystyle{spphys}
\bibliography{bibliography}

\end{document}